\documentclass[showpacs,preprintnumbers,prd,nofootinbib,floats,amssymb,floatfix]{revtex4}
\usepackage{amsmath}
\usepackage{amsfonts}
\usepackage{graphicx}
\usepackage{hyperref}
\usepackage{lipsum}
\usepackage{tikz}
\usepackage{caption}
\usepackage[english]{babel}
\usetikzlibrary{decorations.pathmorphing, arrows.meta, positioning, calc, patterns,angles,quotes}
\usepackage{algorithm}
\usepackage{algorithmicx}
\usepackage{algpseudocode}
\usepackage{comment}
\usepackage{placeins}

\usepackage{amsmath}
\usepackage{amstext}
\usepackage[inline]{trackchanges}
 
\begin{document}

\title{Singular Lagrangians and the Hamilton-Jacobi formalism in classical mechanics}
\author{Luis G. Romero-Hernández$^{\dagger}$, Jaime Manuel-Cabrera$^{\ddagger}$, Ramón E. Chan-López$^{\ast}$, and Jorge M. Paulin-Fuentes$^{\oplus}$}

\email{$^{\dagger}$192a12006@egresados.ujat.mx,$^{\ddagger}$ jaime.manuel@ujat.mx, $^{\ast}$eduardo.clopez13@gmail.com, and $^{\oplus}$jorge.paulin@ujat.mx}

 \affiliation{Divisi\'on Acad\'emica de Ciencias B\'asicas, Universidad Ju\'arez Aut\'onoma de Tabasco,  \\
  Km 1 Carretera
Cunduac\'an-Jalpa, Apartado Postal 24, 86690 Cunduac\'an, Tabasco, M\'exico,}

\begin{abstract}

\begin{center}
    Abstract
\end{center}
\vspace{-1.5mm}

This work conducts a Hamilton–Jacobi analysis of classical dynamical systems with internal constraints. We examine four systems, all previously analyzed by David Brown: three with familiar components (point masses, springs, rods, ropes, and pulleys) and one chosen specifically for its detailed illustration of the Dirac-Bergmann algorithm's logical steps. Including this fourth system allows for a direct and insightful comparison with the Hamilton-Jacobi formalism, thereby deepening our understanding of both methods. To provide a thorough analysis, we classify the systems based on their constraints: non-involutive, involutive, and a combination of both. We then use generalized brackets to ensure the theory's integrability, systematically remove non-involutive constraints, and derive the equations of motion. This approach effectively showcases the Hamilton-Jacobi method's ability to handle complex constraint structures. Additionally, our study includes an analysis of a gauge system, highlighting the versatility and broad applicability of the Hamilton-Jacobi formalism. By comparing our results with those from the Dirac-Bergmann and Faddeev-Jackiw algorithms, we demonstrate that the Hamilton-Jacobi approach is simpler and more efficient in its mathematical operations and offers advantages in computational implementation.

\vspace{0.25cm}

%{Palabras clave: Formalismo de Hamilton-Jacobi, Sistema hamiltoniano constreñido, Sistema no involutivo, Corchetes Generalizados.}


\end{abstract}
 \date{\today}
%\pacs{98.80.-k,98.80.Cq}

\preprint{}
\maketitle

\section{INTRODUCTION}

Singular systems are crucial in physics because the theories that describe fundamental interactions are gauge theories derived from a singular Lagrangian. Due to their significance, various formalisms have been developed to address these singular systems, with the "Dirac-Bergmann Algorithm" being the most extensively studied method in the literature \cite{Dirac01, Dirac02, Dirac03, Dirac, Ber, Ber1}. The Dirac-Bergmann formalism classifies constraints into first-class and second-class types and determines Hamiltonian functions that encompass all constraint information. Dirac’s classification differentiates first-class constraints, which have zero Poisson brackets with other constraints, from second-class constraints, which have non-zero Poisson brackets. For second-class constraints, Dirac introduced a new Poisson bracket, known as the Dirac bracket, corresponding to the commutator. This distinction is considered highly significant by most physicists in both classical and quantum mechanics \cite{Teitelboim, Sun, Hanson}.

However, while elegant, the Dirac-Bergmann formalism is complex, involving many intricate logical steps \cite{Brown1}. This complexity has led to the development of alternative methods for analyzing singular systems. One such method is the Faddeev-Jackiw approach, which simplifies the process by defining a Lagrangian linearized in velocities and using a symplectic structure to derive generalized brackets from a matrix known as the Faddeev-Jackiw symplectic matrix \cite{Faddeev, Garcia}. Extended by Barcelos-Neto and Wotzasek \cite{Barcelos-Neto, Barcelos-Neto2}, this approach offers a different perspective on managing constrained systems.

Another powerful alternative is the Hamilton-Jacobi (HJ) formalism \cite{Guler1}. Originally developed by Carathéodory \cite{Cara} for regular systems with first derivatives, the HJ method was later generalized by Güler to address singular systems \cite{Guler2, Guler3}. This formalism has since been expanded to include Lagrangians with higher-order derivatives \cite{Pi}, Berezinian systems \cite{Pi2}, linear actions \cite{Pi3}, and various applications in gravitational fields \cite{Pi4} and theories involving topologically massive particles \cite{Pi5}.

The Hamilton-Jacobi (HJ) formalism is particularly valuable for exploring the dynamics of constrained systems in classical mechanics and serves as a fundamental tool for understanding their quantum counterparts. This formalism is advantageous across various contexts, including field theory, gauge theories, and general relativity. Although much research has focused on first-class constraints \cite{Gra, Gra1, Esca} and alternatives for second-class constraints have been developed \cite{BFT1, Batalin, Loran, Baleanu}, it is important to acknowledge that the HJ formalism is not confined to these complex scenarios. Indeed, its applicability extends to finite-dimensional systems, which are simpler compared to the intricate field theories with an infinite number of degrees of freedom. For example, in mechanical systems, the HJ formalism has been successfully applied to various models \cite{Bertin2, Bertin, Nawafleh1, Nawafleh2, Na, Rothe}, demonstrating its functionality and relevance across different types of systems.

Our analysis aims to apply the HJ formalism to mechanical systems with a finite number of degrees of freedom, presenting the fundamental concepts clearly and concisely. Notably, the systems we will examine have been previously analyzed using the Dirac-Bergmann formalism in \cite{Brown1, Brown2} and the Faddeev-Jackiw formalism in \cite{Arellano}. Within this context, we revisit the solution for discrete singular systems to further explore their dynamics under the Hamilton-Jacobi formalism. Section II introduces the Hamilton-Jacobi formalism, as discussed in references \cite{Bertin, Bertin2}. Subsequent sections apply this formalism to systems proposed by Brown \cite{Brown1, Brown2}. Section III details the solution of a pendulum suspended from two springs, Section IV describes three masses arranged in a ring connected by springs, and Section V covers a system of masses, springs, and pulleys characterized as involutive. Section VI analyzes Brown’s system \cite{Brown1}, illustrating the logical steps in the Dirac–Bergmann algorithm and their relation to the Hamilton-Jacobi formalism.

The appendix provides detailed computations of the Hamilton-Jacobi algorithm using code implemented in Wolfram Mathematica. This symbolic computation enables an efficient and direct exploration of the method’s possibilities.

\section{THE HAMILTON-JACOBI FORMALISM}

%La mayoría de los siguientes puntos a tratar son expuestos de la misma manera en \cite{Bertin}, e incluso detallados a mayor profundidad. Lo que se muestra a continuación sintetiza lo presentado allí, y muestra directamente la forma de operar bajo el formalismo.

%La mayoría de los puntos que se abordarán a continuación siguen la misma exposición que se encuentra en \cite{Bertin}, y además se profundizan en mayor medida. Lo que se presenta a continuación resume lo expuesto en ese documento, mostrando directamente cómo aplicar el formalismo correspondiente.

Most of the points about to be addressed below follow the same approach presented in \cite{Bertin}. What follows summarizes the content presented there and directly shows how to operate within the formalism.

\subsection{Method considerations}

Consider a system described by $N$ generalized coordinates denoted as $q^{i}$, with their associated velocities $\dot{q}^{i}$, which can be described by the action integral functional
\begin{equation}
    I = \int_{t_{0}}^{t_{1}} L\left( t, q^{i}, \dot{q}^{i} \right) dt.
    \label{1}
\end{equation}

%To achieve an extremal configuration of the above equation, there must be a function $S(q^{i},t)$ ( the Hamilton principal function) that satisfies the following conditions \cite{Cara}

 To achieve an extremal configuration of the equation above, a function  $S(q^{i},t)$, known as the Hamiltonian principal function, must satisfy the following conditions \cite{Cara}

%Para que se tenga una configuración extremal de la ecuación anterior, se debe contar con una función $S(q^{i},t)$ que satisfaga las siguientes condiciones \cite{Cara}:

\begin{eqnarray}\label{2}
    \frac{\partial L}{\partial \dot{q}^{i}} = \frac{\partial S}{\partial q^{i}},
\end{eqnarray}

\begin{eqnarray}\label{3}
    \frac{\partial S}{\partial t} + \frac{\partial S}{\partial q^{i}} \dot{q}^{i} = L.
\end{eqnarray}

%Resolver la ecuación \eqref{3} para $S$ es el sentido del formalismo de Hamilton-Jacobi \cite{Bertin}. Esto resulta una tarea relativamente sencilla para sistemas regulares, en los que tenemos expresiones bien definidas para las velocidades $\dot{q}^{i}$ en términos de las coordenadas y derivadas de $S$ (momentos conjugados). Para aquellos sistemas que sean singulares, es decir, cuya matriz Hessiana

%Solving equation \eqref{3} for $S$ is the essence of the Hamilton-Jacobi formalism \cite{Bertin}. This task is relatively straightforward for regular systems, where well-defined expressions for the velocities $\dot{q}^{i}$ in terms of the coordinates and derivatives of $S$ (conjugate momenta) are available. For systems that are singular, i.e., those whose Hessian matrix is not invertible, the process becomes more complex and requires additional considerations.

%Solving equation \eqref{3} for $S$ is the essence of the Hamilton-Jacobi formalism \cite{Bertin}. This task is relatively straightforward for regular systems, where well-defined expressions for the velocities $\dot{q}^{i}$ in terms of the coordinates and derivatives of $S$ (conjugate momenta) are available. For systems that are singular, i.e., those whose Hessian matrix is undefined or degenerate, additional considerations and methods are required. The method begins by analyzing the Lagrangian as a function of n generalized coordinates.

The HJ formalism is obtained by converting equation \eqref{3} into a partial differential equation for $S$ \cite{Bertin}. This conversion is possible if the velocities $\dot{q}^{i}$ can be expressed in terms of the coordinates and derivatives of $S$. Equation \eqref{2} can be inverted to derive these expressions, assuming the Hessian condition

\begin{eqnarray}
    det (A_{ij})= det \bigg(\frac{\partial^{2}L}{\partial \dot{q}^{i}\partial \dot{q}^{j}}\bigg) \neq 0, \quad i,j=1,...,N,
    \label{4a}
\end{eqnarray}

is satisfied. 

If conditions \eqref{4a} are not  satisfied, one may consider that the Hessian has rank $P\leqslant N$. 
%For systems that are singular, i.e., those whose Hessian matrix is undefined or degenerate, additional considerations and methods are required.
%Entonces existen $R=N-P$ momentos dependientes. Los momentos genaralizados $p_{i}$ correspondientes a las coordenadas generalizadas $q^{i}$ son definidos como 
Therefore, there are $R=N-P$ dependent moments. The generalized moments $p_{i}$ corresponding to the generalized coordinates 
$q^{i}$ are defined as 
 \begin{eqnarray}
     p_{a}&=&\frac{\partial L}{\partial \dot{q}^{a}}, \quad a=1,...,P, \label{4b1} \\
      p_{z}&=&\frac{\partial L}{\partial \dot{q}^{z}}, \quad z=P+1,...,N,
      \label{4b2}
 \end{eqnarray}

%donde $q^{i}$ es divido en dos sets, $q^{a}$ y $q^{z}$. Para las $P$ coordenadas asociadas a la parte regular de la matriz Hessiana, se le denotará con $q^{a}$, mientras que a las $R=N-P$ coordenadas asociadas a la parte nula, se les etiquetará con $t^{z} \equiv q^{z}$. Ya que el rango de la matriz Hessiana \eqref{4a} es $(N-R)$, se pueden despejar las $P$ velocidades  $\dot{q}^{a}$ de (5) como 

where $q^{i}$ is divided into two sets,  $q^{a}$ and $q^{z}$.
The $P$ coordinates associated with the regular part of the Hessian matrix are denoted by $q^{a}$, while the 
$R=N-P$ coordinates associated with the null part are labeled as $t^{z} \equiv q^{z}$. Since the rank of the Hessian matrix \eqref{4a} is $(N-R)$, the 
$P$ velocities  $\dot{q}^{a}$ can be determined from equation \eqref{4b1} as
\begin{eqnarray}\label{4c}
    \dot{q}^{a}=\dot{q}^{a}(q^{i},\dot{t}^{z},p_{a};t).
\end{eqnarray}

%De esta manera adoptamos la misma notación que se tiene en la referencia \cite{Bertin}. Así pues, las ecuaciones en las que se puede despejar las $P$ velocidades $\dot{q}^{a}$ tendrían la forma:

%\begin{equation}\label{4}
 %   \dot{q}^{a} = \phi^{a} \left( t, t^{z}, q^{b}, \frac{\partial S}{\partial q^{b} } \right),
%\end{equation}

%donde: $\mu=P+1,..., N$ y $a,b=1,...,P.$

The remaining equations cannot be inverted, but they must still be valid. We can write them as follows
\begin{equation}\label{5}
    \frac{\partial S}{\partial t^{z}} + H_{z}\left( t^{z} ,q^{a}, \frac{\partial S }{ \partial q^{a} };t\right) = 0, \hspace{1cm}    H_{z} \equiv - \frac{\partial \mathcal{L}}{\partial \dot{t}^{z} }  \Bigg|_{\dot{q} = \phi }.
\end{equation}

The hamiltonian function $H_0$ will have the expected form
\begin{equation}\label{6}
    H_{0} \equiv \frac{\partial S}{\partial t^{z}}\dot{t}^{z} + \frac{\partial S}{ \partial q^{a}} \phi^{a} - L \left(t^{z}, q^{a}, \dot{t}^{z}, \phi^{a};t \right),
\end{equation}

however, this would not depend on $\dot{t}^{z}$ since equation \eqref{5} is not invertible by assumption; it does not depend on this variable. The same happens in \eqref{4c}. Therefore, by evaluating these relations in \eqref{3}, we obtain that
\begin{equation}\label{7}
     \frac{\partial S}{\partial t} + H_{0}\left( t^{z},q^{a}, \frac{\partial S}{\partial q^{a}};t \right) = 0 .
\end{equation}

%This is the Hamilton-Jacobi equation. By defining $p_{0}=\partial_{0}S$ as the canonical momentum corresponding to the variable $x_{0}=t$, these equations can be expressed in a unified form. The equations \eqref{5} and \eqref{7} form a set of equations called Hamilton-Jacobi partial differential equations (HJPDEs), which can be written in a shorthand notation as

This represents the Hamilton-Jacobi equation. By introducing 
$p_{0}=\partial_{0}S$  as the canonical momentum associated with the variable $t^{0}=t$, the equations can be combined into a single, coherent form. To further illustrate this, consider Equations \eqref{5} and \eqref{7}, which form the basis of the Hamilton-Jacobi partial differential equations (HJPDEs). These equations capture the essence of the Hamilton-Jacobi framework and can be succinctly represented as follows

 \begin{equation}\label{8}
      \frac{\partial S}{\partial t^{\alpha}} + H_{\alpha} \left( t^{\beta}, q^{a}, \frac{\partial S}{\partial q^{a}} \right) = 0, \hspace{1cm} \alpha,\beta=0,P+1,...,N.
 \end{equation}

%Where it is defined $t^0 \equiv t$.
%The conjugate momenta are defined, in the context of the Hamilton-Jacobi formalism, as\footnote{This notation adopted for the conjugate momenta using $\pi$ and $p$ changes for the examples shown later. It is used in this context for the description under development.}:

The conjugate momenta are defined in the context of the Hamilton-Jacobi formalism as follows

\begin{equation}\label{9}
     \pi_{\alpha} \equiv \frac{\partial S}{\partial t^{\alpha}}, \hspace{1cm} p_{a} \equiv \frac{\partial S}{\partial q^{a}}.
\end{equation}

%Now we can define the following functions, which, in the first instance, can be identified with the primary constraints in the Dirac formalism:

Now, we can define the following functions, which can initially be identified with the primary constraints in the Dirac formalism

\begin{equation}\label{10}
     H'_{\alpha} (t^{\beta}, q^{a}, \pi_{\beta},p_{a} ) \equiv \pi_{\alpha} + H_{\alpha} (t^{\beta}, q^{a}, p_{a}) = 0.
\end{equation}

%This relation is easily obtained due to the equivalence between \eqref{2} and \eqref{5}, with the first equation of \eqref{9}.

This relation is easily derived from the equivalence between \eqref{2} and \eqref{5} with the first equation of \eqref{9}.

%The independence between variables $t^{\alpha}$ associated to the constraints $H'_{\alpha}$, permits to have total diferential equations of motion:

The independence of the variables $t^{\alpha}$ associated with the constraints $H'_{\alpha}$ allows for the formulation of total differential equations of motion

\begin{equation}\label{11}
	dq^{a} = \frac{\partial H'_{\alpha}}{ \partial p^{a} } dt^{\alpha}, \hspace{1 cm} dp_{a} = - \frac{\partial H'_{\alpha}}{ \partial q^{a} } dt^{\alpha}.
\end{equation}

%De igual manera, el diferencial total de la acción:

Similarly, the total differential of the action is given by

\begin{equation}\label{12}
    dS = p_{a} dq^{a} - H_{\alpha}dt^{\alpha} = \left( -H_{\alpha} + p_{a} \frac{\partial H'_{\alpha}}{\partial p_{a}} \right)dt^{\alpha}.
\end{equation}

%The system defined by equations \eqref{11} and \eqref{12} represents the characteristic equations (CEs) of the HJPDEs. The integrability of these equations, and consequently the independence of the variables, will be confirmed using the Frobenius Integrability Condition, as elaborated in references \cite{Bertin, Guler2}.

The system defined by equations \eqref{11} and \eqref{12} represents the characteristic equations (CEs) of the HJPDEs. To confirm the integrability of these equations and, consequently, the independence of the variables, we will use the Frobenius Integrability Condition, as detailed in references \cite{Bertin, Guler2}.

This condition can be summarized as

\begin{equation}\label{12a}
	dH'_{\alpha} = \left\{ H'_{\alpha}, H'_{\beta} \right\}dt^{\beta} = 0.
\end{equation}

Initially, the integrability of the system is addressed by primary constraints. If primary constraints prove insufficient, secondary constraints must be identified and their integrability verified. This process is repeated iteratively until all possible constraints are identified, ranging from primary to secondary, tertiary, and beyond, resulting in a total of $k$  constraints. The process concludes when at least one Poisson bracket is non-zero, i.e.,  $\left\{ H'_{\alpha}, H'_{\beta} \right\} \neq 0 $.

%Initially, it may be covered by the primary constraints. If not, secondary constraints will be obtained, and their integrability must also be verified, repeating the process until all possible constraints are identified, ranging from primary to secondary, tertiary, and so on, resulting in $k$ constraints. The process ends when more than one Poisson bracket is different from zero $\left\{ H'_{\alpha}, H'_{\beta} \right\} \neq 0 $.

Once all possible constraints have been identified, we will have a final set of $k+1$ HJPDEs in the form \eqref{10}. Some constraints may not be associated with independent variables. Therefore, arbitrary variables must be linked to these new constraints (specifically with the secondary and subsequent constraints, if any), expanding the space of independent variables \cite{Bertin2}.

With all constraints identified and a final set of supposedly independent parameters $t^{\alpha}$ in the extended parameter space, the integrability of the constraints will be reverified, this time using the fundamental differential

\begin{equation}\label{13}
    dF = \left\{ F, H'_{\alpha} \right\} dt^{\alpha}.
\end{equation}

In systems with only primary constraints, the evolution of any quantity or dynamic variable $F$ can still be determined using equation \eqref{13}, regardless of the dimensionality of the independent variable space indexed by  $\alpha$. If the integrability condition \eqref{12a} is satisfied for all identified constraints, the system will be completely integrable and involutive. Conversely, if only a subset of constraints is integrable, the system will be partially integrable \cite{Bertin2}.

%In a system with only primary constraints, the evolution of any quantity or dynamic variable  $F$ could be obtained in the same way by \eqref{13}, regardless of the dimension of the space of independent variables covered by the index  $\alpha$.
%If the integrability condition \eqref{12} for all obtained constraints is satisfied, we will have a completely integrable and involutive system. If only a subset of constraints is integrable, we will have a partially integrable system \cite{Bertin2}.

The principal function, as a solution to the HJPDEs, often simplifies the process of obtaining system trajectories compared to directly solving the equations of motion \eqref{11} or \eqref{13}. Both methods ultimately yield the same dynamics, as demonstrated in \cite{Nawafleh1}. It is important to note that this approach is particularly effective for completely integrable systems, and operates similarly for regular systems. For systems that are not fully involutive, alternative methods are available \cite{BFT1, Batalin, Loran, Baleanu, Heredia}.

%The principal function, as a solution to the HJPDEs, often allows for obtaining these trajectories in a simpler way than solving the system of equations of motion \eqref{11} or \eqref{13}. Either approach yields the same dynamics, as exemplified in \cite{Nawafleh1}. It should be noted that this scheme works very well with completely integrable systems, and in such cases, operates in the same manner as with regular systems. For systems that are not completely involutive, there are alternative methods \cite{BFT1, Batalin, Loran, Baleanu, Heredia}.

To address systems with a final set of non-integrable HJPDEs, the approach involves expressing the fundamental differential \eqref{13} as follows

\begin{equation}\label{17}
    dF = \left[ \left\{ F, H'_{0} \right\} - \left\{ F, H'_{z} \right\} \left( M^{-1} \right)^{zx} \left\{ H'_{x}, H'_{0}\right\} \right] dt,
\end{equation}

where $M_{xz} = \left\{ H'_{x}, H'_{z} \right\} $ is a Poisson bracket (PB) matrix between constraints, and $\left( M^{-1} \right)^{zx}$ is its inverse. It is noteworthy that all independent variables, except for time $t$, were disregarded. 

%The development that entails this relationship, and the subsequent discussion of the Generalized Bracket, are detailed in reference \cite{Bertin2}.

Now, defining the Generalized Bracket as \cite{Bertin2}
\begin{equation}\label{18}
 \left\{ F,G \right\}^{*} \equiv \left\{ F,G \right\}- \left\{ F, H'_{z} \right\} \left( M^{-1} \right)^{zx}\left\{H'_{x}, G \right\},
\end{equation}

equation \eqref{17} is simplified to
\begin{equation}\label{19}
    dF = \left\{ F, H'_{0} \right\}^{*} dt.
\end{equation}

%For this particular case, the matrix $M_{xz}$ has the characteristic of being regular because all of the constraints are non-involutive (second class). \footnote{There can exist cases when non-involution of all constraints does not lead to regularity of the matrix. This will be seen in the last example (Brown's System)}

%In this case, the matrix $M_{xz}$ is regular because all constraints are non-involutive (second class)\footnote{There can be cases where non-involution of all constraints does not lead to matrix regularity. This will be illustrated in the final example (Brown's System).}.

%In this case, the matrix $M_{xz}$ is regular because all constraints are non-involutive (second class). However, there can be instances where the non-involution of all constraints does not result in matrix regularity, which will be demonstrated in the final example (Brown's System).

In this case, the matrix $M_{xz}$ is regular because all constraints are non-involutive (second class). However, there can be instances where the non-involution of all constraints does not result in matrix regularity, which will be analyzed in detail in the final example of this work (Brown's System).

%For partially integrable systems, the process is similar; however, now only those independent variables associated with the subset of non-involutive constraints must be eliminated. In this situation, the matrix $M_{xz}$ will be singular, thus having a rank $r < k$, and therefore $r$ non-involutive constraints. The idea is to get rid of the $r$ independent variables associated with that regular part of the matrix. To do this, a matrix $M_{\Bar{b} \Bar{a}} = \left\{ H'_{\Bar{b}}, H'_{\Bar{a}} \right\}$, with: $\Bar{a}, \Bar{b} = 1,...,r.$ is constructed. This is a Poisson bracket matrix bewteen non-involutive constraints, which is regular. In this way, the fundamental differential takes the following form:
For partially integrable systems, the process is similar; however, only those independent variables associated with the subset of non-involutive constraints must be eliminated. In this situation, the matrix $M_{xz}$ will be singular, having a rank $r < k$ and therefore $r$ non-involutive constraints. The goal is to eliminate the $r$ independent variables associated with the regular part of the matrix. To achieve this, construct a matrix $M_{\Bar{b} \Bar{a}} = \left\{ H'_{\Bar{b}}, H'_{\Bar{a}} \right\}$, where $\Bar{a}, \Bar{b} = 1, \ldots, r$, which is a Poisson bracket matrix between non-involutive constraints and is regular. The fundamental differential then takes the following form

\begin{equation}\label{20}
\begin{split}
     dF &= \left[ \left\{ F, H'_{\Bar{\alpha}} \right\} - \left\{ F, H'_{\Bar{a}} \right\} \left( M^{-1} \right)^{\Bar{a} \Bar{b}} \left\{ H'_{\Bar{b}}, H'_{\Bar{\alpha}} \right\}  \right]dt^{\Bar{\alpha}} \\
     &= \left\{ F, H'_{\Bar{\alpha}} \right\}^{*} dt^{\Bar{\alpha}}.
\end{split}
\end{equation}

Where: $\Bar{\alpha} = 0, r+1,...,k. $ 

%So, the Poisson bracket has the following structure:
Thus, the Poisson bracket takes the following form

\begin{equation}\label{21}
\left\{ F, G \right\}^{*} = \left\{ F, G \right\} - \left\{ F, H'_{\Bar{a}} \right\} \left( M^{-1} \right)^{\Bar{a} \Bar{b}} \left\{ H'_{\Bar{b}}, G \right\}.
\end{equation}

%It should be mentioned that the Generalized Brackets for this case can be defined for any set of non-involutive constraints; that is, it is not necessary to have the final set of obtained constraints. If other missing constraints were to appear, they will do so through the condition \cite{Bertin2}:
It should be noted that the GB can be defined for any set of non-involutive constraints; it is not necessary to have the final set of obtained constraints. If additional missing constraints appear, they will do so through the condition \cite{Bertin2}

\begin{equation}\label{22}
    \left\{ H'_{\Bar{x}}, H'_{0} \right\}^{*} = 0,
\end{equation}

where: $\Bar{x} = r+1,...,k. $

%Thus, we can obtain the differential equations of motion through \eqref{19} or \eqref{20}, as the case may be.
Thus, the differential equations of motion can be obtained from \eqref{19} or \eqref{20}, depending on the case.

%In general, it is possible to count the degrees of freedom of a system with $Q$ dynamic variables in phase space, and with $k$ constraints, of which $r$ are non-involutive and $k-r$ are involutive, as follows:

\subsection{HJ physical degrees of freedom}

As previously mentioned, the HJ approach does not require classifying constraints as first class or second class. Consequently, we can count the physical Degrees Of Freedom (DOF) as follows

%In general, the degrees of freedom of a system with $Q$ dynamic variables in phase space and $k$ constraints, of which $r$ are non-involutive and $k-r$ are involutive, can be calculated as follows:

\begin{equation}\label{DOF}
 \# \: \text{DOF} = \frac{1}{2} \left( Q - 2K - r \right).
\end{equation}

Where $Q$ is the dynamic variables in phase space, $r$ are non-involutive constraints  and $K$ are involutive constraints.

\subsection{Gauge transformations}

%En \cite{Bertin} se muestra como el conjunto completo de restricciones involutivas $H'_{\alpha}=0$ son generadoras de transformaciones cánonicas infinitesimales con la forma

In \cite{Bertin}, it is shown that the complete set of involutive constraints $H'_{\alpha} = 0$ generate infinitesimal canonical transformations of the form\footnote{Where $\xi^{I}=(t^{\alpha},q^{a},\pi_{\alpha},p_{a})$.}

\begin{equation}\label{22a}
    \delta \xi^{I} = \left\{\xi^{I}, H'_{\alpha} \right\}\delta t^{\alpha},
\end{equation}

which are called the characteristic ﬂows  (CFs) of the system. 
%where $\xi^{I}=(t^{\alpha},q^{a},\pi_{\alpha},p_{a})$.
%Entre las CFs (\eqref{22a}), se pueden definir una clase de transformaciones especiales. Estas transformaciones se pueden relacionar con las transformaciones de gauge. 

%Among the CFs \eqref{22a}, a special class of transformations can be defined. These transformations can be connected to gauge transformations. By setting $\delta t^{0}=\delta t=0$, we ensure that the transformations are considered at a fixed moment in time, implying that no changes in time are accounted for during these transformations. Thus, in addition to requiring that $\delta\xi^{I}$ are evaluated at constant $t$, we can impose that they remain within the reduced phase space. In this scenario, equation \eqref{22a} represents the same transformations referred to as "point transformations" by Dirac \cite{Dirac} in the Hamiltonian framework.

Among the characteristic flows (CFs) defined in \eqref{22a}, we can identify a special class of transformations that are related to gauge transformations. By setting $\delta t^{0}=\delta t=0$, we ensure that these transformations are evaluated at a fixed moment in time, meaning that they do not account for temporal changes. Consequently, besides requiring that $\delta\xi^{I}$
be evaluated at constant $t$, we can also impose that they remain within the reduced phase space. Under these conditions, equation \eqref{22a} describes transformations analogous to those referred to as "point transformations" by Dirac \cite{Dirac} in the Hamiltonian framework.

Given a set of involutive constraints $H'_{z}$ associated with the independent variables $t^{z}$, the transformation is given by
\begin{equation}\label{23}
    \delta \xi^{I} = \left\{\xi^{I}, G \right\},
\end{equation}

where $G = H'_{z} \delta t^{z}$ is the generating function. Consequently, the above equation can also be expressed as
\begin{equation}\label{24}
    \delta_{G} \xi^{I} = \left\{\xi^{I}, H'_{z} \right\} \delta t^{z}.
\end{equation}

%These transformations will be Noether symmetries as long as, at constant time, the Lie equation holds 

%An in-depth analysis of the construction of gauge generators through the HJ formalism can be found in \cite{Bertin}

A thorough analysis of the construction of gauge generators through the HJ formalism can be found in \cite{Bertin}.

%These transformations will be Noether symmetries if, at constant time, the Lie equation holds 

%\begin{equation}\label{25}
%    \delta L = \left( \frac{\partial L}{\partial q^{i}} - \frac{d}{dt} \frac{\partial L}{\partial \dot{q}^{i}}  \right)\delta q^{i} +  \frac{d}{dt} \left( \frac{\partial L}{\partial \dot{q}^{i}} \delta q^{i}\right) = 0.
%\end{equation}

%For a more detailed analysis, we refer to \cite{Bertin}.

\section{Pendulum and two springs}

%In this section it is analyzed the second system exposed in \cite{Brown2}, and consists of a mass $m$ attached to a rigid pendulum which length is denoted by $l$, and which in turn  is connected to two springs hanging from a ceiling, separated by a distance of $2d$. The Cartesian coordinates $x$ and $y$ correspond to the point of junction between the two springs and the pendulum, while the angle $\theta$ pertains to the pendulum and its deflection is measured relative to the negative y-axis. The origin is located at the midpoint between the connections of the springs to the ceiling.

This section analyzes the second system described in \cite{Brown2}. The system consists of a mass $m$ attached to a rigid pendulum with a length denoted by $l$. This pendulum is connected to two springs hanging from a ceiling, separated by a distance of $2d$. The Cartesian coordinates $x$ and $y$ correspond to the junction point between the two springs and the pendulum. The angle $\theta$ describes the pendulum's deflection, measured relative to the negative y-axis. The origin is located at the midpoint between the connections of the springs to the ceiling.

%%%%%%%%%%%%%%%%%%%%%%%%%%%%%%%%%%%%%%%%%%%%%%%%%%
\begin{figure}[!ht]
    \begin{center}

\begin{tikzpicture}

% Parámetros
\def\d{2} % Distancia desde el origen a los puntos de anclaje de los resortes
\def\l{2.25} % Longitud del péndulo 
\def\angle{60} % Ángulo del péndulo en grados (30 grados a la derecha del eje Y negativo)

% Coordenadas de los puntos
\coordinate (A) at (-\d,-0.1); % Punto de anclaje izquierdo
\coordinate (B) at (\d,-0.1); % Punto de anclaje derecho
\coordinate (O) at (1,-1.5); % Punto de unión de los resortes
\coordinate (P) at ({\l*sin(\angle)},{-1.5-\l*cos(\angle)}); % Masa del péndulo

% Dibujar la barra superior
\filldraw[fill=gray!50] (-\d-0.2,-0.1) rectangle (\d+0.2,0.1);

% Dibujar los resortes
\draw[thick,decorate,decoration={coil,aspect=0.7,amplitude=4pt,segment length=5pt}] (A) -- (O);
\draw[thick,decorate,decoration={coil,aspect=0.4,amplitude=4pt,segment length=2.5pt}] (B) -- (O);

% Dibujar la barra del péndulo
\draw[thick] (O) -- (P);

% Dibujar la masa m
\filldraw[fill=red] (P) circle (0.2);

% Dibujar la línea punteada vertical
\draw[dashed] (O) -- (1,-2.7);

% Dibujar la línea curva para representar el ángulo
\draw[thin] (1,-2.2) arc [start angle=270, end angle=299, radius=1];

% Etiquetas de los resortes
\node at (-\d*0.5+0.2,-1.05) {$k$};
\node at (\d-0.25,-1) {$k$};

%Etiqueta de la masa
\node at ({\l*sin(\angle) +0.5},{-1.5-\l*cos(\angle)}) {$m$};

%Etiqueta del ángulo
\node at (1.3,-2.4) {$\theta$};

% Dibujar ejes
\draw[->] (\d+0.2,-0.1) -- (\d+0.7,-0.1) node[right] {$x$};
\draw[->] (0,0.1) -- (0,0.7) node[above] {$y$};

\end{tikzpicture}

\end{center}
    \caption{Pendulum and two springs attached to the ceiling at the junction points $x=\pm d$ and $y=0$.}
    \label{fig:PenduloDosResortes}
\end{figure}
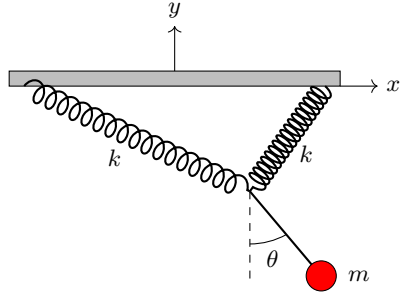 

%%%%%%%%%%%%%%%%%%%%%%%%%%%%%%%%%%%%%%%%%%%%%%%%%%%

%The system analyzed in this section is deﬁned by the Lagrangians:

The system analyzed in this section is defined by the following Lagrangian
	\begin{equation}\label{26}
		L = \frac{m}{2} \left(\dot{x}^{2} + \dot{y}^{2} + l^{2}\dot{\theta}^{2}\right) + ml\left(\dot{x} \cos \theta + \dot{y} \sin \theta\right) \dot{\theta} - mg\left( y - l \cos \theta \right) - k \left(x^{2} + y^{2} + d^{2} \right) .
	\end{equation} 

 %The Lagrangian function \ref{26} is singular, since the rank of the Hessian matrix

%The Lagrangian function \ref{26} is singular due to the rank of the Hessian matrix

The Lagrangian function \eqref{26} is singular because the rank of the Hessian matrix
 
\begin{equation}\label{27}
A_{ij}=\begin{pmatrix}
		m & 0 & ml \cos \theta \\
		0 & m & ml \sin \theta \\
		 ml \cos \theta &  ml \sin \theta & ml^{2}
	\end{pmatrix} ,
\end{equation}

is two. The associated null vector obtained from $A_{ij}$ is
\begin{equation}\label{28}
    (N_{1})_{i} = \left( -l \cos \theta, -l \sin \theta, 1 \right).
\end{equation}

The conjugate momenta are:
 \begin{eqnarray}\label{29}
	p_{x} &=& \frac{\partial L}{\partial \dot{x}} = m \dot{x} + ml\dot{\theta} \cos \theta , \\
	p_{y} &=& \frac{\partial L}{\partial \dot{y}} = m \dot{y} + ml\dot{\theta} \sin \theta , \\
	p_{\theta} &=& \frac{\partial L}{\partial \dot{\theta}} = ml \left(\dot{x} \cos \theta + \dot{y} \sin \theta \right) +ml^2\dot{\theta}.
\end{eqnarray}

%Next, by means of the aforementioned equations of the conjugate momenta, it is necessary to construct a vector $\phi^{i}$ of constraints that have the form: 

Next, using the aforementioned equations for the conjugate momenta, we need to construct a vector $\phi^{i}$ of constraints, which takes the form:

\begin{equation}\label{30}
   \phi^{i} = \begin{pmatrix}
		p_{x} - m \dot{x} - ml\dot{\theta} \cos \theta \\
		p_{y} - m \dot{y} - ml\dot{\theta} \sin \theta  \\
		p_{\theta} - ml \left(\dot{x} \cos \theta + \dot{y} \sin \theta \right) - ml^2\dot{\theta}
	\end{pmatrix}
 =
 \begin{pmatrix}
		0\\
		0\\
		0
	\end{pmatrix},
\end{equation}

%through which we are going to obtain the correct primary constraint $H'_{1}$ (since there is only one null vector, we will have only one primary constraint), which will be found by means of:

through this, we obtain the correct primary constraint, $H'_{1}$. Since there is only one null vector, we will have only one primary constraint, which will be determined by

\begin{equation}\label{31}
    H'_{1} = (N_{1})_{i} \phi^{i} = 0.
\end{equation}

This is
\begin{equation}\label{32}
H'_{1} = p_{\theta} -  lp_{x} \cos \theta - l p_{y} \sin \theta = 0  .
\end{equation}

The subsequent step in developing the HJ formalism involves calculating the canonical Hamiltonian. The canonical Hamiltonian $H_{0}$ is defined by the conventional method of expressing

\begin{equation}\label{32a}
p_{i}\dot{q}_{i}-L(q,\dot{q}),
\end{equation}

in terms of $p's$ and $q's$. Although it is not possible to solve for all $\dot{q}'s$ in terms of  $p's$ and $q's$, it can be demonstrated that the expression \eqref{32a} depends solely on the phase space variables \cite{Dirac,Hanson}. For our specific example, the canonical Hamiltonian is

%Therefore, the canonical hamiltonian is:

\begin{equation}\label{33}
	H_{0} = \frac{\left( p_{x}^{2} + p_{y}^{2} \right)}{2m} + k \left( x^{2} + y^{2} + d^{2} \right) + mg\left(y-l\cos \theta\right) .
\end{equation}

Additionally, the canonical variables adhere to the following Poisson brackets

\begin{eqnarray}
    \left\{x, p_{x}\right\}=1, \quad \left\{y, p_{y}\right\}=1, \quad
    \left\{\theta, p_{\theta}\right\}=1.
\end{eqnarray}

So far, we have the set of HJPDEs
\begin{equation}\label{34}
	\begin{split}
		& H'_{0} = p_{0} + H_{0} = 0 ,\\
		& H'_{1} = p_{\theta} -  lp_{x} \cos \theta - l p_{y} \sin \theta  = 0   .
	\end{split}
\end{equation}

Then, the fundamental differential is written
\begin{equation}\label{35}
	dF  = \left\{F, H'_{0}\right\}dt + \left\{ F, H'_{1} \right\} d\theta .
\end{equation}

Now, verifying the integrability
\begin{equation}\label{36}
	dH'_{1} = \left\{H'_{1},H'_{0}\right\}dt + \left\{H'_{1},H'_{1}\right\}d\theta = 2kl\left(x \cos \theta + y \sin \theta \right) dt.
\end{equation}

%Since $dH'_{1}$ is not identically zero, a secondary constraint is obtained

As $dH'_{1}$ is not identically zero, a secondary constraint can be derived

\begin{equation}\label{37}
	H'_{2} = 2kl \left(x \cos \theta + y \sin \theta \right) = 0 .
\end{equation}

To ensure the system is integrable, we must also verify the variation of 
$H'_{2}$. Through calculations, we determined that 

\begin{equation}\label{38}
	\begin{split}
	    dH'_{2} &= \left\{H'_{2},H'_{0}\right\}dt + \left\{H'_{2},H'_{1}\right\}d\theta ,\\ 
     &= \frac{2kl}{m} \left(p_{x} \cos \theta + p_{y} \sin \theta \right) dt - 2kl \left(l + x \sin \theta - y \cos \theta \right) d\theta.
	\end{split}
\end{equation}

We note that $\left\{H'_{2}, H'_{0}\right\} \neq 0$ and  $\left\{H'_{2}, H'_{1}\right\} \neq 0$, therefore, there will not be any new constraint.

It is apparent that we have a final set of HJPDEs
\begin{equation}\label{39}
	\begin{split}
		& H'_{0} = p_{0} + H_{0} = 0 ,\\
		& H'_{1} = p_{\theta} -  lp_{x} \cos \theta - l p_{y} \sin \theta  = 0  , \\
		& H'_{2} =2kl \left(x \cos \theta + y \sin \theta \right) = 0 .
	\end{split}
\end{equation}

From the constraints $H'_{1}$ and $H'_{2}$, we obtain the relations 
\begin{equation}\label{rel}
	\begin{split}
		& \theta = - \arctan \left(\frac{x}{y}\right), \\
		& p_{\theta} = \frac{l}{\sqrt{x^{2} + y^{2}}} \left(x p_{y} - y p_{x}\right), 
	\end{split}
\end{equation}
\begin{equation}\label{48}
	\begin{split}
		& \sin \theta = \frac{x}{\sqrt{x^{2} + y^{2}}}, \\
		& \cos \theta = - \frac{y}{\sqrt{x^{2} + y^{2}}}, \\
		& r \equiv \sqrt{x^{2} + y^{2}}.
	\end{split}
\end{equation}

We assumed that $|y| = - y$ since $y$ is always a negative quantity, indicating that the pendulum is consistently below the ceiling.
%The relations \eqref{rel}-\eqref{48} can be used freely to manipulate and simplify the results of the Poisson brackets, as will be shown later. These relations are identical to those presented in \cite{Brown2}, where second-class constraints were resolved to derive them.
The relations in equations \eqref{rel} and \eqref{48} will be used to manipulate and simplify the results of the Poisson brackets, as will be demonstrated later. These relations are consistent with those derived in \cite{Brown2}, where second-class constraints were resolved to obtain them.

%Next, the integrability of the constraints will be verified through the extended fundamental differential 
With these relations established, we will next verify the integrability of the constraints using the extended fundamental differential

\begin{equation}\label{49}
	dF = \left\{F, H'_{0}\right\}dt + \left\{ F, H'_{1} \right\} d\theta + \left\{F, H'_{2}\right\}d\omega ,
\end{equation}

where only the independent variable $\omega$ associated with the secondary constraint $H'_{2}$ was added.  

Thus, the evolution of the constraint turns out to be
\begin{equation}\label{50}
	\begin{split}
		& dH'_{1} = \left\{H'_{1}, H'_{0}\right\}dt + \left\{ H'_{1}, H'_{1} \right\} d\theta + \left\{H'_{1}, H'_{2}\right\}d\omega, \\
		&\hspace{6mm} = 2kl\left(x \cos \theta + y \sin \theta \right) dt +2kl \left(l + x \sin \theta - y \cos \theta \right) d\omega  ,
	\end{split}
\end{equation}
\begin{equation}\label{51}
	\begin{split}
		& dH'_{2} = \left\{H'_{2}, H'_{0}\right\}dt + \left\{ H'_{2}, H'_{1} \right\} d\theta + \left\{H'_{2}, H'_{2}\right\}d\omega ,\\
		&\hspace{6mm} = \frac{2kl}{m} \left(p_{x} \cos \theta + p_{y} \sin \theta \right) dt - 2kl \left(l + x \sin \theta - y \cos \theta \right) d\theta .
	\end{split}
\end{equation}

All the constraints are non-involutive, as evidenced by their evolution, which shows that more than one Poisson bracket is nonzero. Therefore, it is necessary to construct the generalized brackets.

%It can be observed that all the constraints are non-involutive, as their evolution results indicate that more than one Poisson bracket is non-zero. Therefore, it is necessary to construct the generalized brackets.

%One can observe that all the constraints are non-involutive, since the results of their evolution show that more than one Poisson bracket is different from zero. It is necessary to construct the Generalized brackets.

The Poisson brackets matrix between constraints $M_{\mu \nu} = \left\{H'_{\mu}, H'_{\nu}\right\}$, with $\mu,\nu = 1,2$ is 
\begin{equation}\label{52}
	M_{\mu \nu}= 2kl \left(l + x \sin \theta - y \cos \theta \right)
	\begin{pmatrix}
		0 & 1 \\
		-1 & 0 
	\end{pmatrix},
\end{equation}

and its inverse
\begin{equation}\label{53}
	\left(M^{-1}\right)^{\nu\mu}= \frac{1}{2kl \left(l + x \sin \theta - y \cos \theta \right)}
	\begin{pmatrix}
		0 & -1 \\
		1 & 0 
	\end{pmatrix}.
\end{equation}

In this way, the GB can be written as
\begin{equation}\label{54}
	 \left\{F, G\right\}^{*} = \left\{F, G\right\} - \left\{F, H'_{1}\right\}\left(M^{-1}\right)^{12} \left\{H'_{2}, G\right\} - \left\{F, H'_{2}\right\}\left(M^{-1}\right)^{21} \left\{H'_{1}, G\right\} .
 \end{equation}

One can also obtain the following nonzero brackets
\begin{equation}\label{55}
	\left\{x, p_{x}\right\}^{*} = 1 - \frac{l \cos^2 \theta}{l + x \sin \theta - y \cos \theta} ,
\end{equation}
\begin{equation}\label{56}
	\left\{y, p_{y}\right\}^{*} = 1 - \frac{l \sin^2 \theta}{l + x \sin \theta - y \cos \theta} ,
\end{equation}
\begin{equation}\label{57}
	\left\{x, p_{y}\right\}^{*} = \left\{y, p_{x}\right\}^{*} =- \frac{l \sin \theta \cos \theta}{l + x \sin \theta - y \cos \theta} ,
\end{equation}
\begin{equation}\label{58}
	\left\{\theta, p_{x}\right\}^{*} = \frac{1}{x \tan \theta + l \sec \theta - y} ,
\end{equation}
\begin{equation}\label{59}
	\left\{\theta, p_{y}\right\}^{*} = \frac{1}{l \csc \theta + x - y \cot \theta} ,
\end{equation}

which through the relations \eqref{48}, can be written as
\begin{equation}\label{60}
	\left\{x, p_{x}\right\}^{*} = \frac{r + lx^{2}/r^{2}}{r + l} ,
\end{equation}
\begin{equation}\label{61}
	\left\{y, p_{y}\right\}^{*} = \frac{r + ly^{2}/r^{2}}{r + l} ,
\end{equation}
\begin{equation}\label{62}
	\left\{x, p_{y}\right\}^{*} = \left\{y, p_{x}\right\}^{*} = \frac{lxy/r^{2}}{r + l} ,
\end{equation}
\begin{equation}\label{63}
	\left\{\theta, p_{x}\right\}^{*} = -\frac{y}{r\left(r + l\right)} ,
\end{equation}
\begin{equation}\label{64}
	\left\{\theta, p_{y}\right\}^{*} = \frac{x}{r\left(r + l\right)} .
\end{equation}

Finally, the characteristic equations written in this same terms turn out to be
\begin{equation}\label{65}
	dx = \left\{x, H'_{0}\right\}^{*} dt = \frac{lx\left(x p_{x} + y p_{y}\right) + p_{x} r^{3}}{mr^{2}\left(r + l\right)} dt ,
\end{equation}
\begin{equation}\label{66}
	dy = \left\{y, H'_{0}\right\}^{*} dt = \frac{ly\left(x p_{x} + y p_{y}\right) + p_{y} r^{3}}{mr^{2}\left(r + l\right)} dt ,
\end{equation}
\begin{equation}\label{67}
	dp_{x} = \left\{p_{x}, H'_{0}\right\}^{*} dt = -2kx \, dt ,
\end{equation}
\begin{equation}\label{68}
	dp_{y} = \left\{p_{y}, H'_{0}\right\}^{*} dt = \left(-2ky - mg\right) dt .
\end{equation}

Hence, the equations of motion are
\begin{equation}\label{69}
	\dot{x} = \left\{x, H'_{0}\right\}^{*} = \frac{lx\left(x p_{x} + y p_{y}\right) + p_{x} r^{3}}{mr^{2}\left(r + l\right)} ,
\end{equation}
\begin{equation}\label{70}
	\dot{y} = \left\{y, H'_{0}\right\}^{*} = \frac{ly\left(x p_{x} + y p_{y}\right) + p_{y} r^{3}}{mr^{2}\left(r + l\right)} ,
\end{equation}
\begin{equation}\label{71}
	\dot{p}_{x} = \left\{p_{x}, H'_{0}\right\}^{*} = -2kx ,
\end{equation}
\begin{equation}\label{72}
	\dot{p}_{y} = \left\{p_{y}, H'_{0}\right\}^{*} = -2ky - mg .
\end{equation}

\vspace{2mm}
The solution to this system of ordinary differential equations (ODEs) is numerical. Once this is obtained, the solution for $\theta(t)$ y $p_{\theta}(t)$ is given directly by the relations \eqref{rel}.

An important feature of any example like \eqref{26} is that the configuration space, represented by the q's, must be at least three-dimensional. This is because the physical degrees of freedom are calculated by taking the dimension of the configuration space, subtracting the number of involutive constraints, and then subtracting half the number of non-involutive constraints. For the example to possess at least one physical degree of freedom, it must include at least one involutive constraint and at least two non-involutive constraints (with the non-involutive constraints always being even in number). Therefore, the configuration space must be at least three-dimensional. 

By determining the constraints and applying \eqref{DOF}, we can establish the system's degrees of freedom  as follows

\begin{equation*}\label{73}
 \# \: \text{DOF} = \frac{1}{2} \left[6 -2(0) - 2 \right] = 2,
\end{equation*}

The system has six dynamical variables, represented by $(x,y,\theta,p_{x},p_{y},p_{\theta})$, and two non-involutive constraints $(H'_{1},H'_{2})$. Thus, the theory exhibits two physical degrees of freedom.

%There are six dynamical variables represented by $(x,y,\theta,p_{x},p_{y},p_{\theta})$, and two non-involutive constraints $(H'_{1},H'_{2})$. Thus, the theory exhibits two physical degrees of freedom.

\newpage
\section{Masses, springs and ring}

%The following system consist of three identical masses $m$ connected by three springs of elastic constant $k$, attached to a ring of radius $R$,  which slide frictionless over the circumference. The origin is located at the center of the ring, and the coordinates $x$ and $y$ correspond to the point of junction bewtween the three springs. The angles $\theta_{1}$, $\theta_{2}$ and $\theta_{3}$, related to each mass, are measured respect to the positive part of the $x$-axis (See ref. \cite{Brown2}).

The following system consists of three identical masses 
$m$ connected by three springs with an elastic constant 
$k$. These springs are attached to a ring of radius $R$, which slides frictionlessly over the circumference. The origin is located at the center of the ring, and the coordinates 
$x$ and $y$ correspond to the junction point of the three springs. The angles $\theta_{1}$, $\theta_{2}$ and $\theta_{3}$, corresponding to each mass, are measured with respect to the positive $x$-axis (see ref. \cite{Brown2}).

%%%%%%%%%%%%%%%%%%%%%%%%%%%%%%%%%%%%%%%%%%%%%%%%%%%%%%%%

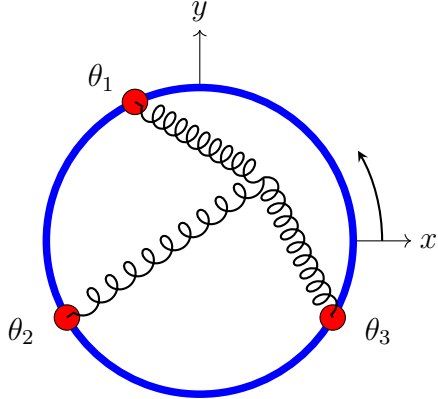
\begin{figure}[!ht]
    \begin{center}

% Inicio de la imagen TikZ
\begin{tikzpicture}
\begin{scope}[shift={(0,0)}, scale=1.2] % Cambia el valor de scale para ajustar el tamaño

% Radio del anillo
\def\R{1.7}

% Punto de unión de los resortes
\coordinate (C) at (0.7, 0.7);

% Ángulos para las masas
\def\angleone{115}  % 90 + 25 = 115
\def\angletwo{210}
\def\anglethree{330}

% Anillo
%\draw[very thick, draw=black] (0,0) circle (\R);
\draw[line width=1mm, color=blue] (0,0) circle (\R);

% Masas
\filldraw[fill=red] (\angleone:\R) circle (4pt);
\filldraw[fill=red] (\angletwo:\R) circle (4pt);
\filldraw[fill=red] (\anglethree:\R) circle (4pt);

% Etiquetas de los ángulos
\node at (\angleone:\R+0.3) [left] {\large $\theta_1$};
\node at (\angletwo:\R+0.3) [left] {\large $\theta_2$};
\node at (\anglethree:\R+0.3) [right] {\large $\theta_3$};

% Resortes
\draw[thick,decorate,decoration={coil,aspect=0.7,amplitude=4pt,segment length=5pt}] (C) -- (\angleone:\R);
\draw[thick,decorate,decoration={coil,aspect=0.9,amplitude=4pt,segment length=8.5pt}] (C) -- (\angletwo:\R);
\draw[thick,decorate,decoration={coil,aspect=0.7,amplitude=4pt,segment length=5.5pt}] (C) -- (\anglethree:\R);

% Ejes coordenados
\draw[->] (\R+0.04,0) -- (\R+0.64,0) node[right] {\large $x$};
\draw[->] (0,\R+0.04) -- (0,\R+0.64) node[above] {\large $y$};

% Flecha curva. Dirección de medición de los ángulos.
\draw[->,>=stealth,thick] (\R+0.32,0) arc[start angle=0,end angle=30,radius=2];

\end{scope}
\end{tikzpicture}
% Fin de la imagen TikZ

\end{center}
    \caption{Three masses sliding without friction in a circular ring, connected by three springs, one attached to each mass.}
    \label{fig:Anillo}
\end{figure} 

%%%%%%%%%%%%%%%%%%%%%%%%%%%%%%%%%%%%%%%%%%%%%%%%%%%%%%%%%%

The Lagrangian that corresponds to this system is
\begin{equation}\label{74}
	L = \frac{mR^{2}}{2} \left(\dot{\theta}_{1}^{2} + \dot{\theta}_{2}^{2} + \dot{\theta}_{3}^{2} \right) - V\left(\theta, x, y\right) .
\end{equation}

With the potential energy $V$
\begin{eqnarray}\label{75a}
	V &=& \frac{k}{2} [\left(x - R \cos \theta_{1}\right)^{2} + \left(y - R \sin \theta_{1}\right)^{2} + \left(x - R \cos \theta_{2}\right)^{2} + \left(y - R \sin \theta_{2}\right)^{2} \nonumber\\ &\,&+ \left(x - R \cos \theta_{3}\right)^{2} +\left(y - R \sin \theta_{3}\right)^{2} ] .
\end{eqnarray}

The hessian matrix obtained from this Lagrangian is
\begin{equation}\label{76}
A_{ij}=\begin{pmatrix}
		0 & 0 & 0 & 0 & 0 \\
		0 & 0 & 0 & 0 & 0 \\ 
		0 & 0 & mR^{2} & 0 & 0 \\
		0 & 0 & 0 & mR^{2} & 0 \\
		0 & 0 & 0 & 0 & mR^{2} \\
	\end{pmatrix} ,
\end{equation}

whose rank is three.

The conjugate momenta are
\begin{eqnarray}\label{77}
    p_{i} &=& \frac{\partial L}{\partial \dot{\theta}_{i}} = mR^{2} \dot{\theta}_{i} , \quad i = 1,2,3,  \\
    p_{x} &=& \frac{\partial L}{\partial \dot{x}} = 0 ,\\
    p_{y} &=& \frac{\partial L}{\partial \dot{y}} = 0 .
\end{eqnarray}

%where $i = 1,2,3$.  

These equations for the conjugate momenta exhibit directly the two  primary constraints $H'_{1} = p_{x} = 0 $ and
$H'_{2} = p_{y} = 0 $.

%Furthermore, the canonical Hamiltonian is given by 

Additionally, the canonical Hamiltonian can be expressed as

\begin{equation}\label{78}
    	H_{0} = \frac{p_{1}^{2} + p_{2}^{2} + p_{3}^{2}}{2mR^{2}} + V\left(\theta,x,y\right),
\end{equation}

%and the canonical variables satisfy the following Poisson brackets

and the canonical variables fulfill the following Poisson bracket relation
\begin{eqnarray}
    \left\{x, p_{x}\right\}=1, \quad \left\{y, p_{y}\right\}=1, \quad
    \left\{\theta_{i}, p_{j}\right\}=\delta_{ij}, \quad i,j=1,2,3.
\end{eqnarray}

In this scenario, there are three HJ equations
\begin{eqnarray}
		 H'_{0} &=& p_{0} + H_{0} = 0 , \label{78a}\\
		 H'_{1} &=& p_{x} = 0, \label{78b}\\
		 H'_{2} &=& p_{y} = 0 \label{78c} .
\end{eqnarray}

We need to examine the integrability of this set. The evolution of any dynamical function F is given by
\begin{equation}\label{80}
	dF = \left\{F, H'_{0}\right\} dt +\left\{F, H'_{1}\right\} dx + \left\{F, H'_{2}\right\} dy .
\end{equation}

Using this differential, we test the integrability of the constraints $(H'_{1},H'_{2})$. Evolving the primary constraints we get
\begin{equation}\label{81}
	\begin{split}
		& dH'_{1} = \left\{H'_{1},H'_{0}\right\}dt + \left\{H'_{1},H'_{1}\right\}dx + \left\{H'_{1},H'_{2}\right\}dy, \\
		& \hspace{6mm} = \left[kR\left(\cos \theta_{1} + \cos \theta_{2} + \cos \theta_{3}\right)- 3kx\right]dt,
	\end{split}
\end{equation}
\begin{equation}\label{82}
	\begin{split}
		& dH'_{2} = \left\{H'_{2},H'_{0}\right\}dt + \left\{H'_{2},H'_{1}\right\}dx + \left\{H'_{2},H'_{2}\right\}dy, \\
		& \hspace{6mm} = \left[kR\left(\sin \theta_{1} + \sin \theta_{2} + \sin \theta_{3}\right)- 3ky\right]dt .
	\end{split}
\end{equation}

Given that $dH'_{1}$ and $H'_{2}$ are not identically zero, the following secondary constraints can be derived
\begin{eqnarray}
		 H'_{3} &=& kR\left(\cos \theta_{1} + \cos \theta_{2} + \cos \theta_{3}\right)- 3kx = 0, \label{83} \\
		 H'_{4} &=& kR\left(\sin \theta_{1} + \sin \theta_{2} + \sin \theta_{3}\right)- 3ky = 0 \label{84}.
\end{eqnarray}

Also verifying the integrability 
\begin{equation}\label{85}
	\begin{split}
		& 	dH'_{3} = \left\{H'_{3},H'_{0}\right\} dt + \left\{H'_{3},H'_{1}\right\} dx + \left\{H'_{3},H'_{2}\right\} dy ,  \\
		& \hspace{6mm} = -\frac{k}{mR} \left(p_{1} \sin \theta_{1} + p_{2} \sin \theta_{2} + p_{3} \sin \theta_{3}\right) dt - 3k dx ,
	\end{split}
\end{equation}
\begin{equation}\label{86}
	\begin{split}
		& 	dH'_{4} = \left\{H'_{4},H'_{0}\right\} dt + \left\{H'_{4},H'_{1}\right\} dx + \left\{H'_{4},H'_{2}\right\} dy , \\
		& \hspace{6mm} = \frac{k}{mR} \left(p_{1} \cos \theta_{1} + p_{2} \cos \theta_{2} + p_{3} \cos \theta_{3}\right) dt - 3k dy .
	\end{split}
\end{equation}

%Since in both expressions, there is more than one Poisson bracket that doesn't vanish.

The integrability of 
$H'_{3}$ and $H'_{4}$ does not yield any new constraints, indicating that the system is non-involutive.

Now, we have a final set of HJPDEs
\begin{equation}\label{87}
	\begin{split}
		& H'_{0} = p_{0} + H_{0} = 0 ,\\
		& H'_{1} = p_{x} = 0 ,\\
		& H'_{2} = p_{y} = 0 ,\\
		& H'_{3} = kR\left(\cos \theta_{1} + \cos \theta_{2} + \cos \theta_{3}\right)- 3kx = 0 ,\\
		& H'_{4} = kR\left(\sin \theta_{1} + \sin \theta_{2} + \sin \theta_{3}\right)- 3ky = 0 .
	\end{split}
\end{equation}

The fundamental differential is now defined as follows
\begin{equation}\label{75}
	dF = \left\{F, H'_{0}\right\} dt +\left\{F, H'_{1}\right\} dx + \left\{F, H'_{2}\right\} dy + \left\{F, H'_{3}\right\} d\tau + \left\{F, H'_{4}\right\} d\lambda .
\end{equation}

%Where the independent parameters $\tau$ and $\lambda$ are related to the constraints $H'_{3}$ and $H'_{4}$ respectively.

%If the integrability of the constraints is verified through \eqref{75}, it will be revealed that all of them are non-involutive, therefor, we procceed to calculate the Poisson bracket matrix between for the construction of the Generalized Bracket. 

The independent parameters  $\tau$ and $\lambda$ are associated with the constraints $H'_{3}$ and $H'_{4}$, respectively. If the integrability of the constraints is verified through \eqref{75}, it will be revealed that all of them are non-involutive. Therefore, we proceed to calculate the Poisson bracket matrix for the construction of the Generalized Bracket.

The Poisson bracket matrix between constraints $M_{\mu \nu} = \left\{H'_{\mu}, H'_{\nu}\right\}$ with $\left\{\mu, \nu\right\} = \left\{1,2,3,4\right\}$ is
\begin{equation}\label{88}
	M_{\mu \nu} = 
	\begin{pmatrix}
		 0 & 0 & 3k & 0 \\
		 0 & 0 & 0 & 3k \\
		 -3k & 0 & 0 & 0 \\
		 0 & -3k & 0 & 0 
	\end{pmatrix} .
\end{equation}

Its inverse is
\begin{equation}\label{89}
	\left(M^{-1}\right)^{\nu \mu} =  \frac{1}{3k}
	\begin{pmatrix}
		0 & 0 & -1 & 0 \\
		0 & 0 & 0 & -1 \\
		1 & 0 & 0 & 0 \\
		0 & 1 & 0 & 0 
	\end{pmatrix}.
\end{equation}

This regular matrix permits that the GB has the structure from \eqref{18}, and then, the fundamental differential as in \eqref{19}. 
%It is worthy to point out that from \eqref{83} and \eqref{84} one can deduct tha
It is worth noting that from \eqref{83} and \eqref{84}, one can deduce that

\begin{eqnarray}
	x &=& \frac{R}{3} \left( \cos \theta_{1} + \cos \theta_{2} + \cos \theta_{3} \right),\label{92}\\
	y &=& \frac{R}{3} \left( \sin \theta_{1} + \sin \theta_{2} + \sin \theta_{3} \right).\label{93}
\end{eqnarray}

Finally, the following non-zero GBs are obtained ($h,k = 1,2,3.$)
\begin{eqnarray}\label{91}
	\left\{\theta_{h}, p_{k}\right\}^{*} &=& \delta_{hk},\\
	\left\{x, p_{h}\right\}^{*} &=& - \frac{R}{3} \sin \theta_{h},\\
	\left\{y, p_{h}\right\}^{*} &=& \frac{R}{3} \cos \theta_{h}.
\end{eqnarray}

The characteristic equations are responsible for governing the evolution of the canonical variables in phase space. From \eqref{75} we obtain the CEs
\begin{equation}\label{92a}
	d \theta_{h} = \left\{\theta_{h}, H'_{0}\right\}^{*} dt = \frac{p_{h}}{mR^2} dt ,
\end{equation}
\begin{equation}\label{93a}
\begin{split}
	d p_{1} = \left\{p_{1}, H'_{0}\right\}^{*}dt = &-\frac{kR}{3} \sin \theta_{1} \left[-3x + R \left( \cos \theta_{1} + \cos \theta_{2} + \cos \theta_{3} \right)\right] dt \\
	 & + \frac{kR}{3} \cos \theta_{1} \left[-3y + R \left( \sin \theta_{1} + \sin \theta_{2} + \sin \theta_{3} \right)  \right] dt \\
     & + kR \left(y \cos \theta_{1} - x \sin \theta_{1}\right)dt,
\end{split}
\end{equation}

\begin{equation}\label{94}
\begin{split}
	d p_{2} = \left\{p_{2}, H'_{0}\right\}^{*}dt = & -\frac{kR}{3} \sin \theta_{2} \left[-3x + R \left( \cos \theta_{1} + \cos \theta_{2} + \cos \theta_{3} \right)\right] dt  \\
	&+ \frac{kR}{3} \cos \theta_{2} \left[-3y + R \left( \sin \theta_{1} + \sin \theta_{2} + \sin \theta_{3} \right)  \right] dt \\
    &+ kR \left(y \cos \theta_{2} - x \sin \theta_{2}\right)dt,
\end{split}
\end{equation}
\begin{equation}\label{95}
\begin{split}
	d p_{3} = \left\{p_{3}, H'_{0}\right\}^{*}dt= & -\frac{kR}{3} \sin \theta_{3} \left[-3x + R \left( \cos \theta_{1} + \cos \theta_{2} + \cos \theta_{3} \right)\right] dt \\
	&+ \frac{kR}{3} \cos \theta_{3} \left[-3y + R \left( \sin \theta_{1} + \sin \theta_{2} + \sin \theta_{3} \right)  \right] dt \\
    &+ kR \left(y \cos \theta_{3} - x \sin \theta_{3}\right)dt .
\end{split}
\end{equation}

After including the relations from \eqref{92} and \eqref{93}, these three results simplify to
\begin{eqnarray}\label{96}
	d p_{1} &=& \frac{kR^2}{3} \left(\sin \left(\theta_{2} - \theta_{1}\right) + \sin \left(\theta_{3} - \theta_{1} \right) \right) dt,\\
	d p_{2} &=& \frac{kR^2}{3} \left(\sin \left(\theta_{3} - \theta_{2}\right) + \sin \left(\theta_{1} - \theta_{2} \right) \right) dt,\\
	d p_{3} &=&  \frac{kR^2}{3} \left(\sin \left(\theta_{1} - \theta_{3}\right) + \sin \left(\theta_{2} - \theta_{3} \right) \right) dt.
\end{eqnarray}

In this way, the equations of motion are:
\begin{eqnarray} 
	\dot{\theta}_{h} &=& \frac{p_{h}}{mR^2}, \label{90a}\\
	\dot{p}_{1} &=& \frac{kR^2}{3} \left(\sin \left(\theta_{2} - \theta_{1}\right) + \sin \left(\theta_{3} - \theta_{1} \right) \right),\label{90b}\\
		\dot{p}_{2} &=& \frac{kR^2}{3} \left(\sin \left(\theta_{3} - \theta_{2}\right) + \sin \left(\theta_{1} - \theta_{2} \right) \right),\label{90c}\\
	\dot{p}_{3} &=& \frac{kR^2}{3} \left(\sin \left(\theta_{1} - \theta_{3}\right) + \sin \left(\theta_{2} - \theta_{3} \right) \right). \label{90d}
\end{eqnarray}

This EDOs system is solved numerically.
Note that \eqref{90a} obtained through the GB, agrees with \eqref{77}. The count of degrees of freedom turns out to be
\begin{equation*}
 \# \: \text{DOF} = \frac{1}{2} \left[ 10 - 2(0) - 4 \right] = 3.
\end{equation*}
The system has ten dynamical variables, represented by $(x,y,\theta_{1},\theta_{2},\theta_{3},p_{x},p_{y},p_{1},p_{2},p_{3})$, and four non-involutive constraints $(H'_{1},H'_{2},H'_{3},H'_{4})$. Thus, the theory exhibits three physical degrees of freedom. It is important to note that the results presented \eqref{90a}, \eqref{90b}, \eqref{90c} and \eqref{90d}, are consistent with those obtained in \cite{Brown2}, which were previously analyzed using the Dirac formalism.

\newpage
\section{Pairs of pulleys}

%The present mechanical system, consists on three pairs of massless pulleys, located one above the other. The upper pulleys are each tied to a mass that in turn is connected to a spring, while the lower ones are fixed. Three ropes that are joined together form a single cord that runs underneath and above, respectively, all the pulleys. (See ref. \cite{Brown2}) 

The current mechanical system consists of three pairs of massless pulleys arranged one above the other. The upper pulleys are each attached to a mass that is connected to a spring, while the lower pulleys are fixed. Three ropes, joined together to form a single cord, run beneath and above all the pulleys. (See ref. \cite{Brown2})

%%%%%%%%%%%%%%%%%%%%%%%%%%%%%%%%%%%%%%%%%%%%%%%%%%%%%%%%%

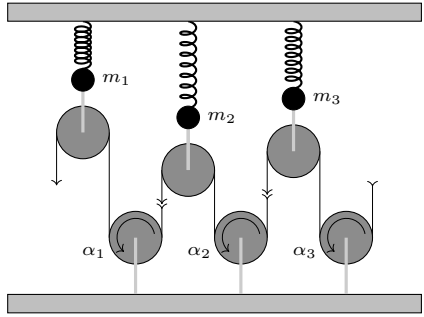
\begin{figure}[!ht]
 \begin{center}
 \begin{tikzpicture}

    \def\r{0.35}
    \coordinate (A) at (-2,-0.12);
    \coordinate (B) at (-2+4*\r,-0.12);
       
    %Resortes
    \draw[thick,decorate,decoration={coil,aspect=0.5,amplitude=3pt,segment length=2pt}] (A) -- (-2,-0.8);
    \draw[thick,decorate,decoration={coil,aspect=0.5,amplitude=3pt,segment length=4pt}] (B) -- (-2+4*\r,-1.3);
    \draw[thick,decorate,decoration={coil,aspect=0.5,amplitude=3pt,segment length=3pt}] (-2+8*\r,-0.12) -- (-2+8*\r,-1.05);

    %Barras de arriba y abajo
    \filldraw[fill=gray!50] (-3,-0.12) rectangle (2.5,0.12);
    \filldraw[fill=gray!50] (-3,-4) rectangle (2.5,-4+0.245);

    %correas
    \draw[ultra thin] (-2+0.35,-1.6) -- (-2+0.35,-3) ;
    \draw[ultra thin] (-2+5*\r,-2.1) -- (-2+5*\r,-3) ;
    \draw[ultra thin] (-2+9*\r,-1.85) -- (-2+9*\r,-3) ;
    \draw[->,ultra thin] (-2-0.35,-1.6) -- (-2-0.35,-2.3) ;
    \draw[->,ultra thin] (-2+3*\r,-2.1) -- (-2+3*\r,-2.1-0.45) ;
    \draw[>-,ultra thin] (-2+3*\r,-2.1-0.45) -- (-2+3*\r,-3) ;
    \draw[->,ultra thin] (-2+7*\r,-1.85) -- (-2+7*\r,-1.85-0.575) ;
    \draw[>-,ultra thin] (-2+7*\r,-1.85-0.575) -- (-2+7*\r,-3) ;
    \draw[>-,ultra thin] (-2+11*\r,-1.85-0.4) -- (-2+11*\r,-3) ;
   
    %Poleas
    \filldraw[fill=gray!90] (-2,-1.6) circle (0.35);
    \filldraw[fill=gray!90] (-2+2*\r,-3) circle (0.35);
    \filldraw[fill=gray!90] (-2+4*\r,-2.1) circle (0.35);
    \filldraw[fill=gray!90] (-2+6*\r,-3) circle (0.35);
    \filldraw[fill=gray!90] (-2+8*\r,-1.85) circle (0.35);
    \filldraw[fill=gray!90] (-2+10*\r,-3) circle (0.35);

    %sujetadores de las poleas
    \draw[very thick, gray!40] (-2,-0.9) -- (-2,-1.6);
    \draw[very thick, gray!40] (-2+2*\r,-3) -- (-2+2*\r,-3.75);
    \draw[very thick, gray!40] (-2+6*\r,-3) -- (-2+6*\r,-3.75);
     \draw[very thick, gray!40] (-2+10*\r,-3) -- (-2+10*\r,-3.75);
    \draw[very thick, gray!40] (-2+4*\r,-1.4) -- (-2+4*\r,-2.1);
    \draw[very thick, gray!40] (-2+4*\r,-1.4) -- (-2+4*\r,-2.1);
    \draw[very thick, gray!40] (-2+8*\r,-1.15) -- (-2+8*\r,-1.85);
    
     %Masas
    \filldraw[fill=black] (-2,-0.9) circle (0.15);
    \filldraw[fill=black] (-2+4*\r,-1.4) circle (0.15);
    \filldraw[fill=black] (-2+8*\r,-1.15) circle (0.15);

    %etiquetas
    \node at (-1.55,-0.9) {\scriptsize $m_{1}$};
    \node at (-2.05+4*\r +0.5,-1.4) {\scriptsize $m_{2}$};
    \node at (-2.05+8*\r +0.5,-1.15) {\scriptsize $m_{3}$};
    \node at (-2+2*\r -0.55,-3.2) {\scriptsize $\alpha_{1}$};
    \node at (-2+6*\r -0.55,-3.2) {\scriptsize $\alpha_{2}$};
    \node at (-2+10*\r -0.55,-3.2) {\scriptsize $\alpha_{3}$};
   
    %Flechas de angulos
    \draw[->,very thin] (-2+2*\r+0.25,-2.95) arc[start angle=10,end angle=230,radius=0.25];
    \draw[->,very thin] (-2+6*\r+0.25,-2.95) arc[start angle=10,end angle=230,radius=0.25];
    \draw[->,very thin] (-2+10*\r+0.25,-2.95) arc[start angle=10,end angle=230,radius=0.25];

      \end{tikzpicture}
  \end{center}
  \caption{Three pair of massless pulleys. The upper ones are attached to the ceiling through springs and masses. A loop of cord winds through all of the pulleys and gets cut at the end of both sides.}
    \label{fig:ParesDePoleas}
\end{figure}

%%%%%%%%%%%%%%%%%%%%%%%%%%%%%%%%%%%%%%%%%%%%%%%%%%%%%%%%%

%The generalized coordinates used to analyze this system are the angles $\alpha_{1}$, $\alpha_{2}$ and $\alpha_{3}$, for the lower pulleys, and they are measured with respect the $x$-axis. The three masses are identical $(m_{1} = m_{2} = m_{3})$, in such a way that $m$ represents any of them.

The generalized coordinates used to analyze this system are the angles $\alpha_{1}$, $\alpha_{2}$ and $\alpha_{3}$ for the lower pulleys, measured with respect to the $x$-axis. The three masses are identical  $(m_{1} = m_{2} = m_{3})$, so $m$ represents any of them.

%Reporting the description given in \cite{Brown2}, to define the kinetic energy, one may note that if the angle $\alpha_{1}$ is incremented by $\delta \alpha_{1}$, the height of the mass $m_{2}$ is incremented by a quantity $R \, \delta \alpha_{1}/2$, where $R$ is the radius of the lower pulley. On the contrary, if the angle $\alpha_{2}$ is incremented by $\delta \alpha_{2}$, the height of the mass $m_{2}$ decreases by a quantity given by $R \, \delta \alpha_{2}/2$. Therefore, the height of the mass $m_{2}$ is $h_{2} = R \left(\alpha_{1}-\alpha_{2}\right) + c$, being $c$ a constant. The same applies for the two remaining masses, so the height of the mass $m_{1}$ is $h_{1}=R \left(\alpha_{3}-\alpha_{1}\right) + c $, and the height of the mass $m_{3}$ is $h_{3}=R \left(\alpha_{2}-\alpha_{3}\right) + c $. Then, the kinetic energy is written as: $T = \left(m/2\right)\left(\dot{h}_{1}^{2} + \dot{h}_{2}^{2} + \dot{h}_{3}^{2}\right)$, or in angular coordinates terms:

According to the description provided in \cite{Brown2}, to define the kinetic energy, one can observe the following: if the angle $\alpha_{1}$ is increased by 
$\delta \alpha_{1}$, the height of the mass $m_{2}$
increases by $R \, \delta \alpha_{1}/2$, where 
$R$ is the radius of the lower pulley. Conversely, if the angle $\alpha_{2}$ is increased by $\delta \alpha_{2}$, the height of the mass $m_{2}$ decreases by $R \, \delta \alpha_{2}/2$. Therefore, the height of the mass $m_{2}$ is given by 
$h_{2} = R \left(\alpha_{1}-\alpha_{2}\right) + c$, where $c$ is a constant. The same applies to the remaining two masses: the height of the mass $m_{1}$
is  $h_{1}=R \left(\alpha_{3}-\alpha_{1}\right) + c $, and the height of the mass $m_{3}$ is $h_{3}=R \left(\alpha_{2}-\alpha_{3}\right) + c $
. Consequently, the kinetic energy is expressed as 
$T = \left(m/2\right)\left(\dot{h}_{1}^{2} + \dot{h}_{2}^{2} + \dot{h}_{3}^{2}\right)$, or in terms of angular coordinates

\begin{equation}\label{97}
	T = \frac{m R^{2}}{8} \left[ \left(\dot{\alpha}_{1} - \dot{\alpha}_{2}\right)^{2} + \left(\dot{\alpha}_{2} - \dot{\alpha}_{3}\right)^{2} + \left(\dot{\alpha}_{3} - \dot{\alpha}_{1}\right)^{2} \right] .
\end{equation}
%The gravitational potential energy $mg\left(h_{1} + h_{2} + h_{3}\right)$ is constant. The spring potential energy is $\left(k/2\right)\left[\left(a - h_{1}\right)^{2} + \left(a-h_{2}\right)^{2} + \left(a-h_{3}\right)^{2}\right]$, where $a$ is a constant that depends on the height of the ceiling y the relaxed length of every spring. Therefore, the potential energy is expressed by
The gravitational potential energy, $mg\left(h_{1} + h_{2} + h_{3}\right)$ remains constant. he spring potential energy is given by $\left(k/2\right)\left[\left(a - h_{1}\right)^{2} + \left(a-h_{2}\right)^{2} + \left(a-h_{3}\right)^{2}\right]$, where $a$ is a constant depending on the height of the ceiling and the relaxed length of each spring. Therefore, the total potential energy is expressed as
\begin{equation}\label{98}
	V\left(\alpha\right) = \frac{kR^{2}}{8} \left[ \left(\alpha_{1} - \alpha_{2}\right)^{2} + \left(\alpha_{2} - \alpha_{3}\right)^{2} + \left(\alpha_{3} - \alpha_{1}\right)^{2}\right].
\end{equation}

The Lagrangian for this system is 

\begin{eqnarray}\label{98a}
    L=\frac{m R^{2}}{8} \left[ \left(\dot{\alpha}_{1} - \dot{\alpha}_{2}\right)^{2} + \left(\dot{\alpha}_{2} - \dot{\alpha}_{3}\right)^{2} + \left(\dot{\alpha}_{3} - \dot{\alpha}_{1}\right)^{2} \right] - V\left(\alpha\right).
\end{eqnarray}

%Thus, it is evident that the Lagrangian is simply the difference $L = T - V$.

The Hessian matrix $A_{ij}$, with $i,j=1,2,3,$ for this Lagrangian is 
\begin{equation}\label{99}
	A_{ij} = \frac{mR^2}{2}
	\begin{pmatrix}
		1 &- \frac{1}{2} &- \frac{1}{2} \\
		-\frac{1}{2} & 1 &- \frac{1}{2} \\ 
	-\frac{1}{2} & -\frac{1}{2} & 1  \\
	\end{pmatrix} ,
\end{equation}
%which rank is two. The only null vector obtained from it is
whose rank is two. The only null vector derived from it is
\begin{equation}\label{100}
    (N_{1})_{i} = (1,1,1).
\end{equation}

The conjugate momenta are
\begin{eqnarray}\label{101}
	p_1 &=& \frac{\partial L}{\partial \dot{\alpha}_{1} } = \frac{mR^2}{4} \left(2\dot{\alpha}_{1} - \dot{\alpha}_{2} - \dot{\alpha}_{3} \right) ,\\
	p_2 &=& \frac{\partial L}{\partial \dot{\alpha}_{2} } = \frac{mR^2}{4} \left(2\dot{\alpha}_{2} - \dot{\alpha}_{1} - \dot{\alpha}_{3} \right) ,\\
	p_3 &=& \frac{\partial L}{\partial \dot{\alpha}_{3} } = \frac{mR^2}{4} \left(2\dot{\alpha}_{3} - \dot{\alpha}_{1} - \dot{\alpha}_{2} \right) .
\end{eqnarray}

The vector $\phi^{i}$, which contains the constraints in the form $f(q,p)=0$ is
\begin{equation}\label{102}
     \phi^{i} = \begin{pmatrix}
		p_{1} - \frac{mR^2}{4} \left(2\dot{\alpha}_{1} - \dot{\alpha}_{2} - \dot{\alpha}_{3} \right)  \\
		p_{2} - \frac{mR^2}{4} \left(2\dot{\alpha}_{2} - \dot{\alpha}_{1} - \dot{\alpha}_{3} \right)  \\
		p_{3} - \frac{mR^2}{4} \left(2\dot{\alpha}_{3} - \dot{\alpha}_{1} - \dot{\alpha}_{2} \right) 
	\end{pmatrix}
 =
 \begin{pmatrix}
		0\\
		0\\
		0
	\end{pmatrix}.
\end{equation}

%so that the correct primary constraint is obtained by means of 
Thus, the correct primary constraint is obtained through
%In this way, the only primary constraint that is obtained is
%In this way, the only primary constraint obtained is
%The  HJ algorithm yields a primary constraint 
\begin{equation}\label{103}
	H'_{1} = (N_{1})_{i} \cdot \phi^{i} = p_{1} + p_{2} + p_{3} = 0 .
\end{equation}

The canonical Hamiltonian $H_{0} = H_{0}(p_1 , p_2, \alpha_1, \alpha_2, \alpha_3)$ is written as 
\begin{equation}\label{104}
	H_{0} = \frac{4}{3 mR^{2}} \left(p_{1} ^{2} + p_{1} p_{2} + p_{2}^{2}\right) + V(\alpha) .
\end{equation}

and the canonical variables satisfy the following Poisson bracket relation
\begin{eqnarray}\label{105a}
\left\{\alpha_{i}, p_{j}\right\} =\delta_{ij}, \quad i,j=1,2,3.
\end{eqnarray}

For now, we have the set of HJPDEs
\begin{equation} \label{106}	
\begin{split}
	H'_{0} & = p_{0} + H_{0} = 0 , \\
	H'_{1} &= p_{1} + p_{2} + p_{3} = 0 .
\end{split}
\end{equation}

%Now, we can know the form of the fundamental differential

%We can now determine the form of the fundamental differential

At this point, we can determine the structure of the fundamental differential

\begin{equation}\label{107}	
	dF = \left\{F, H'_{0}\right\} dt + \left\{F, H'_{1}\right\} d\alpha_{3} .
\end{equation}

%Then, verifying the integrability of $H'_{1}$

Next, by verifying the integrability of $H'_{1}$

\begin{equation}\label{108}	
\begin{split}
		dH'_{1} &= \left\{H'_{1}, H'_{0}\right\} dt + \left\{H'_{1}, H'_{1}\right\} d\alpha_{3} = 0. \\
\end{split}
\end{equation}

%it is clear that the integrability condition for $H'_{1}$ is identically satisfied.
%and $H'_{1}$ has its integrability identically satisﬁed.
%This means that the set of equations \eqref{102} is completely integrable. The algebra  $ \left\{H'_{1}, H'_{0}\right\}$ and  $\left\{H'_{1}, H'_{1}\right\}$ indicate that the set $(H'_{0},H'_{1})$ is involutive with the PBs, it is clear that the integrability condition for $H'_{1}$ is identically satisfied.
This implies that the set of equations \eqref{102} is fully integrable. The algebra $\left\{H'_{1}, H'_{0}\right\}$ and $\left\{H'_{1}, H'_{1}\right\}$ demonstrate that the set $(H'_{0},H'_{1})$ is involutive with respect to the Poisson brackets, ensuring that the integrability condition for $H'_{1}$ is inherently satisfied.
%These equations can be written explicitly as:
%\begin{equation} \label{105}
%    H'_{0} = \frac{\partial S}{\partial t} +  \frac{4}{3 mR^{2}} \left[ \left( \frac{\partial S}{\partial \alpha_{1}}\right) ^{2} +\left( \frac{\partial S}{\partial \alpha_{1}}\right) \left( \frac{\partial S}{\partial \alpha_{2}}\right) + \left( \frac{\partial S}{\partial \alpha_{2}}\right)^{2}\right] + V(\alpha) = 0,
%\end{equation}
%\begin{equation}
   % H'_{1} = \frac{\partial S}{\partial \alpha_{1}} + \frac{\partial S}{\partial \alpha_{2}} + \frac{\partial S}{\partial \alpha_{3}} = 0.
%\end{equation}

%Note that \eqref{105} cannot be solved by separation of variables.

%Now, the characteristic equations are

Using \eqref{107}, one find the CEs
\begin{equation}\label{109}	
	d\alpha_{1} = \frac{4}{3mR^2} \left( 2 p_{1} + p_{2} \right) dt + d\alpha_{3},
\end{equation}
\begin{equation}\label{110}	
	d\alpha_{2} = \frac{4}{3mR^2} \left( 2 p_{2} + p_{1} \right) dt + d\alpha_{3},
\end{equation}
\begin{equation}\label{111}	
	d\alpha_{3} =  d\alpha_{3},
\end{equation}
\begin{equation}\label{112}	
	dp_{1} = \frac{kR^{2}}{4} \left( -2\alpha_{1} + \alpha_{2} + \alpha_{3} \right) dt,
\end{equation}
\begin{equation}\label{113}	
	dp_{2} = \frac{kR^{2}}{4} \left( -2\alpha_{2} + \alpha_{1} + \alpha_{3} \right) dt,
\end{equation}
\begin{equation}\label{114}	
	dp_{3} = - dp_{2} - dp_{1} =\frac{kR^{2}}{4} \left( -2\alpha_{3} + \alpha_{1} + \alpha_{2} \right) dt.
\end{equation}

%Or, alluding to the independence among the variables $\alpha_{3}$ and $t$

%Alternatively, considering the independence of the variables

Since integrability is assured, 
$\alpha_{3}$ and  $t$ are independent variables. Consequently, the time evolution leads to the following set of equations
\begin{equation}\label{115}	
	\dot{\alpha}_{1} = \frac{4}{3mR^2} \left( 2 p_{1} + p_{2} \right),
\end{equation}
\begin{equation}\label{116}	
	\dot{\alpha}_{2} = \frac{4}{3mR^2} \left( 2 p_{2} + p_{1} \right),
\end{equation}
\begin{equation}\label{117}	
	\dot{p}_{1} = \frac{kR^{2}}{4} \left( -2\alpha_{1} + \alpha_{2} + \alpha_{3} \right),
\end{equation}
\begin{equation}\label{118}	
	\dot{p}_{2} = \frac{kR^{2}}{4} \left( -2\alpha_{2} + \alpha_{1} + \alpha_{3} \right),
\end{equation}
\begin{equation}\label{119}	
	\dot{p}_{3} = -\dot{p}_{2} - \dot{p}_{1} =\frac{kR^{2}}{4} \left( -2\alpha_{3} + \alpha_{1} + \alpha_{2} \right). 
\end{equation}

The canonical transformations of the theory are derived by setting $dt = 0$ in the characteristic equations.  Consequently, the generator of these transformations is the involutive Hamiltonian $H'_{1}$. 
The Hamiltonian $H'_{1}$ generates the canonical transformations given by %\footnote{$t^{1}$ is the independent variable associated to $H'_{1}$, therefore it is equal to $\alpha_{3}$.} 

\begin{eqnarray}\label{120}	
    \delta F=\{F,H_{1}\}\delta t^{1}, 
\end{eqnarray}

where $t^{1}$, corresponding to  $H'_{1}$, is equivalent to $\alpha_{3}$. The transformations that arise are
\begin{eqnarray}\label{121}	
    \delta \alpha_{i}=\delta t^{1}, \quad \delta p_{i}=0, \quad i=1,2,3.
\end{eqnarray}

%the gauge transformations $\delta \xi = \left\{\xi,H'_{1} \right\} \delta t^{1}$ are obtained as follows
%\begin{eqnarray}\label{118}
%	\delta \alpha_{i} &=& \delta t^{1}, \\
%	\delta p_{i} &=& 0.
%\end{eqnarray}
%Where $i=1,2,3$, and $t^{1}$ is the independent variable associated to $H'_{1}$, therefore it is equal to $\alpha_{3}$. 
The transformation of $\alpha_{i}$ is arbitrary, depending on the specific form of $\delta t^{1}$. The generator of the characteristic flows (CF) along the direction of $t^{1}$ is expressed as $ G_{CF}=H_{1}\delta t^{1}$.

% \begin{eqnarray}\label{122}	
 %    G_{CF}=H_{1}\delta t^{1}.
 %\end{eqnarray}
%The transformations \eqref{118} are Noether symmetries by the fact that they keep the Lagrangian invariant
Finally, the first variation of equation \eqref{98a} under infinitesimal transformations $\delta \alpha_{i}=\alpha'_{i}-\alpha_{i}$ is given by 
\begin{equation}\label{123}	
	\begin{split}
        &\delta L = \\
		&\left[ \frac{kR^{2}}{4} \left(-2\alpha_{1} + \alpha_{2}+ \alpha_{3} \right)- \frac{mR^{2}}{4} \left( 2 \ddot{\alpha}_{1} - \ddot{\alpha}_{2} - \ddot{\alpha}_{3} \right) \right] \delta \alpha_{1} + \frac{d}{dt} \left[\frac{mR^{2}}{4} \left( 2 \dot{\alpha}_{1} - \dot{\alpha}_{2} - \dot{\alpha}_{3} \right)\delta \alpha_{1} \right] \\
		&+\left[ \frac{kR^{2}}{4} \left(-2\alpha_{2} + \alpha_{1}+ \alpha_{3} \right)- \frac{mR^{2}}{4} \left( 2 \ddot{\alpha}_{2} - \ddot{\alpha}_{1} - \ddot{\alpha}_{3} \right) \right] \delta \alpha_{2} + \frac{d}{dt} \left[\frac{mR^{2}}{4} \left( 2 \dot{\alpha}_{2} - \dot{\alpha}_{1} - \dot{\alpha}_{3} \right)\delta \alpha_{2} \right] \\
		&+\left[ \frac{kR^{2}}{4} \left(-2\alpha_{3} + \alpha_{1}+ \alpha_{2} \right)- \frac{mR^{2}}{4} \left( 2 \ddot{\alpha}_{3} - \ddot{\alpha}_{1} - \ddot{\alpha}_{2} \right) \right] \delta \alpha_{3} + \frac{d}{dt} \left[\frac{mR^{2}}{4} \left( 2 \dot{\alpha}_{3} - \dot{\alpha}_{1} - \dot{\alpha}_{2} \right)\delta \alpha_{3} \right]. \\ 
	\end{split}
\end{equation}
%If \eqref{121} are symmetries of the Lagrangian function, $\delta L$ must be zero. Then, substituting \eqref{121} in \eqref{123}, we have
If the expressions in \eqref{121} are symmetries of the Lagrangian function,  $\delta L$ must be zero. Then, substituting \eqref{121} in \eqref{123}, we have
\begin{eqnarray}\label{124}
    \delta L = (p_{1}+p_{2}+p_{3})\frac{d }{dt}(\delta t^{1}).
\end{eqnarray}

In this scenario, $\delta L= 0$ when $\delta t^{1}$ is constant. Assuming  $\delta t^{1}=\epsilon$, the generator gauge is 
\begin{eqnarray}\label{125}
    G_{g} = \epsilon H'_{1}. 
\end{eqnarray}

The corresponding gauge transformations are
\begin{eqnarray}\label{126}
    \delta{\alpha_{i}}=\{\alpha_{i},H'_{1}\}=\epsilon, \nonumber\\
    \delta{p_{i}}=\{p_{i},H'_{1}\}=0.
\end{eqnarray}

%Note that \eqref{126}, which is derived from \eqref{125}, is consistent with the results reported in \cite{Brown2}. The number of degrees of freedom is given by
%\begin{equation*}
 %\# \: \text{DOF} = \frac{1}{2} \left[ 6 - 2(1) - 0 \right] = 2. 
%\end{equation*}

%The system has six dynamical variables, represented by $(\alpha_{1},\alpha_{2},\alpha_{3},p_{1},p_{2},p_{3})$, and one involutive constraints $H'_{1}$. Thus, the theory exhibits two physical degrees of freedom. It is important to note that the results presented \eqref{115}-\eqref{119} and \eqref{126}, are consistent with those obtained in \cite{Brown2}, which were previously analyzed using the Dirac formalism.

%Note that \eqref{126}, derived from \eqref{125}, is consistent with the results reported in \cite{Brown2}. The number of degrees of freedom is calculated as follows

It is important to note that equation \eqref{126}, which was derived from equation \eqref{125}, is consistent with the results reported in \cite{Brown2}. After confirming this consistency, we proceed to calculate the number of degrees of freedom using \eqref{DOF}

\begin{equation*}\label{127}
\# \text{DOF} = \frac{1}{2} \left[ 6 - 2(1) - 0 \right] = 2.
\end{equation*}

%The system involves six dynamical variables, denoted by $(\alpha_{1}, \alpha_{2}, \alpha_{3}, p_{1}, p_{2}, p_{3})$, and one involutive constraint, $H'_{1}$. Therefore, the theory exhibits two physical degrees of freedom. It is important to emphasize that the results presented in \eqref{115}-\eqref{119} and \eqref{126} are consistent with those obtained in \cite{Brown2}, which were previously analyzed using Dirac's formalism.
The system consists of six dynamic variables, represented as $(\alpha_{1}, \alpha_{2}, \alpha_{3}, p_{1}, p_{2}, p_{3})$, and a single involutive constraint, $H'_{1}$. Consequently, the theory has two physical degrees of freedom. It is noteworthy that the results shown in \eqref{115}-\eqref{119} and \eqref{126} align with those found in \cite{Brown2}, which were previously examined using Dirac's formalism.

\section{Brown's system}

Brown \cite{Brown1} introduced a model designed to provide a thorough example that encompasses all the steps of the Dirac-Bergmann formalism. This model is crucial because many current examples only cover a few specific steps, leaving students to piece together a full understanding of the algorithm on their own. Brown’s model offers a comprehensive view, making it easier to grasp each step and how all components are related.
Given the intricate nature of the Dirac-Bergmann algorithm, which involves a complex series of logical steps, using a more streamlined method can be beneficial. The Hamilton-Jacobi approach presents a valuable alternative in this context. By applying the Hamilton-Jacobi framework to Brown's model, we aim to offer a more integrated and accessible perspective, demonstrating how this method can enhance understanding and simplify the analysis in comparison to the Dirac formalism.

The system discussed in this section is characterized by the Lagrangian
%The singular Lagrangian is
\begin{equation}\label{128}
	L = \frac{1}{2} \left[\left(q_{1} + \dot{q}_{2} + \dot{q}_{3}\right)^{2} + \left(\dot{q}_{4} - \dot{q}_{2}\right)^{2} + \left(q_{1} + 2q_{2}\right) \left(q_{1} + 2q_{4}\right) \right] .
\end{equation}

%Its hessian matrix
The matrix of second derivatives of $L$ with respect to the
velocities $\dot{q}_{i}$
\begin{equation}\label{129}
	A_{ij} = 
	\begin{pmatrix}
		0 & 0 & 0 & 0 \\
		0 & 2 & 1 & -1 \\ 
		0 & 1 & 1 & 0 \\
		0 & -1 & 0 & 1
	\end{pmatrix} .
\end{equation}
%is singular with rank two and the null vectors associated
%The system is singular, with a matrix rank of two. Additionally, it has the following null vectors
The system is singular with a rank of two. Additionally, it has the following null vectors
%\begin{eqnarray}
 %   (N_{1})_{i} &=&(0,1,-1,1),  \\
  %  (N_{2})_{i} &=& (1,0,0,0).
%\end{eqnarray}
\begin{eqnarray}
    (N_{1})_{i} &=&(1,0,0,0),  \\
    (N_{2})_{i} &=& (0,1,-1,1).
\end{eqnarray}

The conjugate momenta are
\begin{eqnarray}\label{130}
	p_{1} &=& \frac{\partial L}{\partial \dot{q}_{1} } = 0 , \\
	p_{2} &=& \frac{\partial L}{\partial \dot{q}_{1} } = q_{1} + 2 \dot{q}_{2} + \dot{q}_{3} - q_{4} ,\\
	p_{3} &=& \frac{\partial L}{\partial \dot{q}_{3} } = q_{1} + \dot{q}_{2} + \dot{q}_{3} , \\
	p_{4} &=& \frac{\partial L}{\partial \dot{q}_{4} } = \dot{q}_{4} - \dot{q}_{2} .
\end{eqnarray}

By means of the above, the vector $\phi^{i}$ of constraints is
\begin{equation}\label{131}
     \phi^{i} = \begin{pmatrix}
		p_{1} \\
		p_{2} - q_{1} - 2 \dot{q}_{2} - \dot{q}_{3} + q_{4} \\
		p_{3} - q_{1} - \dot{q}_{2} - \dot{q}_{3} \\
        p_{4} - \dot{q}_{4} + \dot{q}_{2}
	\end{pmatrix}
 =
 \begin{pmatrix}
		0\\
		0\\
		0\\
        0
	\end{pmatrix}.
\end{equation}

The correct primary constraints are obtained by
\begin{eqnarray}\label{132}
    H'_{1} &=& (N_{1})_{i} \cdot \phi^{i} = p_{1} ,\\
    H'_{2} &=& (N_{2})_{i} \cdot \phi^{i} = p_{2} - p_{3} + p_{4}=0.
\end{eqnarray}

%This is
%\begin{eqnarray}
%	H'_{1} &=& p_{2} - p_{3} + p_{4} = 0 , \\
%	H'_{2} &=& p_{1} = 0 .
%\end{eqnarray}

The canonical Hamiltonian can be written as
\begin{equation}\label{133}
 H_{0} = \frac{1}{2} \left[p_{3}^{2} + p_{4}^{2} - 2 p_{3}q_{1} - \left(q_{1} + 2q_{2}\right)\left(q_{1} + 2q_{4}\right) \right] .
\end{equation}

Up to now, we have a first set of HJPDEs 
\begin{equation}\label{134}
\begin{split}
	H'_{0} &= p_{0} + H_{0} = 0 , \\
	H'_{1} &= p_{1} = 0 ,\\
	H'_{2} &=  p_{2} - p_{3} + p_{4} = 0 .	
\end{split}
\end{equation}

%Where $p_{0} = \partial S / \partial t$.

The canonical variables satisfy the Poisson bracket relation $\{q_{i},p_{j}\}=\delta_{ij}$ $(i,j=1,2,3,4)$. We have verified that the PBs between the expressions in \eqref{134} are $ \left\{H'_{1},H'_{1}\right\}=0$, $\left\{H'_{1},H'_{2}\right\}$, $ \left\{H'_{1},H'_{0}\right\}=- 2(q_{1} + q_{2} + q_{4})$ and $\left\{H'_{2},H'_{0}\right\}=(p_{3} + q_{1} + q_{2} + q_{4})$. As a result, the system of HJ equations is not integrable, necessitating the imposition of an additional constraint
\begin{equation}\label{135}
	H'_{3} = 2(q_{1} + q_{2} + q_{4}) = 0 ,
\end{equation}
\begin{equation}\label{136}
	H'_{4} = p_{3} + q_{1} + q_{2} + q_{4} = 0 .
\end{equation}

Therefore, the fundamental differential is given by
\begin{equation}\label{137}
	dF = \left\{F,H'_{0}\right\} dt + \left\{F,H'_{1}\right\}dq_{1} + \left\{F,H'_{2}\right\}dq_{2}.
\end{equation}

%The evolution of the constraints results in being
%\begin{equation}
%\begin{split}
%		dH'_{1}& = \left\{H'_{1},H'_{0}\right\} dt + \left\{H'_{1},H'_{1}\right\}dq_{2} + \left\{H'_{1},H'_{2}\right\}dq_{1} = 0 \\
%		& = 2(q_{1} + q_{2} + q_{4})dt ,
%\end{split}
%\end{equation}
%\begin{equation}
%	\begin{split}
%		dH'_{2}& = \left\{H'_{2},H'_{0}\right\} dt + \left\{H'_{2},H'_{1}\right\}dq_{2} + \left\{H'_{2},H'_{2}\right\}dq_{1} = 0 \\
%		&= (p_{3} + q_{1} + q_{2} + q_{4}) dt .
%	\end{split}
%\end{equation}

%Hence, two secondary constraints are obtained
%\begin{equation}
%	H'_{3} = 2(q_{1} + q_{2} + q_{4}) = 0 ,
%\end{equation}
%\begin{equation}
%	H'_{4} = p_{3} + q_{1} + q_{2} + q_{4} = 0 .
%\end{equation}

Now verifying the integrability of \eqref{135} and \eqref{136}
\begin{equation}\label{138}
	\begin{split}
		dH'_{3}& = \left\{H'_{3},H'_{0}\right\} dt + \left\{H'_{3},H'_{1}\right\}dq_{1} + \left\{H'_{3},H'_{2}\right\}dq_{2}  \\
		&= 2p_{4} dt + 4dq_{1} + 2dq_{2} ,
	\end{split}
\end{equation}
\begin{equation}\label{139}
	\begin{split}
		dH'_{4}& = \left\{H'_{4},H'_{0}\right\} dt + \left\{H'_{4},H'_{1}\right\}dq_{1} + \left\{H'_{4},H'_{2}\right\}dq_{2}  \\
		&= p_{4} dt + 2dq_{1} + dq_{2} .
	\end{split}
\end{equation}

%This indicates that no further constraints will be obtained. We now have a final set of HJPDEs
This indicates that no additional constraints will be derived. We now arrive at the final set of HJPDEs
\begin{equation}\label{140}
\begin{split}
	H'_{0} &= p_{0} + H_{0} = 0 , \\
	H'_{1} &= p_{1} = 0 ,\\
	H'_{2} &= p_{2} - p_{3} + p_{4} = 0 , \\
	H'_{3} & = 2(q_{1} + q_{2} + q_{4}) = 0 , \\
	H'_{4} & = p_{3} + q_{1} + q_{2} + q_{4} = 0 .
\end{split}
\end{equation}

The fundamental differential now takes the form
\begin{equation}\label{141}
	dF = \left\{F,H'_{0}\right\} dt + \left\{F,H'_{1}\right\}dq_{1} + \left\{F,H'_{2}\right\}dq_{2} + \left\{F,H'_{3}\right\} d\phi + \left\{F,H'_{4}\right\} d\psi .
\end{equation}

%Where $\phi$ and $\psi$ are the independent variables associated to the constraints $H'_{3}$ and $H'_{4}$ respectively.
%Now, it is necessary to evolve all the constraints

Here, $\phi$ and $\psi$ represent the independent variables associated with the constraints $H'_{3}$ and $H'_{4}$, respectively.
%It is now necessary to evolve all the constraints.
%The integrability conditions $dH'_{\alpha}$ assumed the form \footnote{where H'_{\alpha} are the canonical contraints \eqref{140}}

At this point, all constraints have been determined and collectively constitute a complete set of Hamilton-Jacobi partial differential equations (HJPDEs), represented in \eqref{140}, corresponding to a set of parameters $t^{\alpha}$. 
%Now, it is necessary to evolve all the constraints \eqref{140}. 

The integrability conditions for $dH'_{\alpha}$ are expressed in the following form 
\begin{eqnarray}\label{142}
   \{H'_{\alpha},H'_{\beta}\}dt^{\beta} =M_{\alpha\beta}dt^{\beta}=0, \quad \alpha,\beta=0,1,2,3,4.
\end{eqnarray}

Where $t^{\alpha}=(t^{0}=t,t^{1}=q_{1},t^{2}=q_{2}, t^{3}=\phi, t^{4}=\psi)$.

By separating the time variable from the other parameters, the integrability conditions specified in \eqref{142} are simplified as follows
\begin{eqnarray}\label{143}
    dH'_{\alpha}=\{H'_{\alpha},H'_{0}\}dt+\{H_{\alpha},H'_{j}\}dt^{j}=0, \quad j=1,2,3,4. 
\end{eqnarray}

First, we analyze the conditions for $H'_{i}$
\begin{eqnarray}\label{144}
    \{H'_{i},H'_{0}\}dt+\{H_{i},H'_{j}\}dt^{j}=0.
\end{eqnarray}

We can introduce an antisymmetric matrix derived from the Poisson brackets between the constraints $H'_{i}$, $M_{ij}=\{H_{i},H'_{j}\}$. Subsequently, it is possible to express

\begin{eqnarray}\label{145}
 M_{ij}dt^{j}=-\{H'_{i},H'_{0}\}dt. 
\end{eqnarray}

If the complete system of HJPDE is not in involution, the M matrix we work with is typically regular. However, consider the scenario where the M matrix is singular with a rank $k<r$. In such a situation, the matrix's singularity implies that there will be a set of $k$ non-involutive constraints. 

To further explore this, let us consider a specific subset of constraints, denoted by $H'_{i}=(H'_{1},H'_{2},H'_{3},H'_{4})$ . Then, the matrix $M_{ij}$ can be written as

\begin{equation}\label{146}
	M_{ij} = 
	\begin{pmatrix}
		0 & 0 & -4 & -2 \\
		0 & 0 & -2 & -1 \\ 
		4 & 2 & 0 & 0 \\
		2 & 1 & 0 & 0
	\end{pmatrix} .
\end{equation}

%The matrix is singular; its rank is equal to two. It is important to mention what this implies.  Es importante hacer mención que esto último implica: Even observing that all the constraints are non-involutive, the matrix $M_{ij}$ is not regular as one could expect. This fact indicates that there is an dependence between constraints.

%The matrix is singular with a rank of two. This characteristic is significant, as it implies that even if all constraints are non-involutive, the matrix $M_{ij}$ does not exhibit regularity as might be expected. This irregularity indicates an underlying dependence between the constraints.

%The matrix is singular with a rank of two. This implies that, despite all constraints being non-involutive, the matrix $M_{ij}$ does not exhibit the regularity one might expect. This lack of regularity suggests an underlying dependence between the constraints.

%The matrix is singular with a rank of two. This implies that there are two non-involutive constraints. As a result of its nullity, the matrix has a nullity of $4−2=2$. Consequently, there are two linearly independent eigenvectors associated with the eigenvalue of zero. The null vector of the matrix \eqref{146} are

The matrix is singular with a rank of two, indicating the presence of two non-involutive constraints. Its nullity, calculated as $4-2=2$ reveals that the matrix has two linearly independent eigenvectors corresponding to the eigenvalue of zero. To illustrate this further, the null vectors of the matrix \eqref{146} are as follows
\begin{eqnarray}
	(\tilde{N_{1}})_{i} &=& (-1, 2, 0, 0),\\
	(\tilde{N_{2}})_{i} &=& (0, 0, -1, 2).
\end{eqnarray}

These allow to obtain a new set of constraints 
\begin{equation*}
	\begin{split}
	&(\tilde{N_{1}})_{i}\cdot H'_{i}=2H'_{2} - H'_{1}  = 2p_{1} - p_{2} + p_{3} - p_{4} = 0, \\
	&(\tilde{N_{2}})_{i}\cdot H'_{i}=2H'_{4}-H'_{3}  = 2(p_{3} + q_{1} + q_{2} + q_{4}) -  2(q_{1} + q_{2} + q_{4}) = 0.
	\end{split}
\end{equation*}

Which can also be written as
\begin{equation}\label{248}
	\begin{split}
	&	H''_{1} = p_{1} - \frac{1}{2} (p_{2} - p_{3} + p_{4}) = 0, \\
	& H''_{2} = p_{3} = 0.
	\end{split}
\end{equation}

%A different combination of constraints gives
We consider the two remaining linear combinations of constraints to be

\begin{equation}\label{249}
\begin{split}
	& 	H''_{3} = \frac{H'_{1} + H'_{2}}{3} = \frac{1}{3} (p_{1} + p_{2} - p_{3} + p_{4}) = 0, \\
	& H''_{4} =  \frac{H'_{3} + H'_{4}}{3} = \frac{1}{3} p_{3} + q_{1} + q_{2} + q_{4} = 0. 
\end{split}
\end{equation}

As discussed earlier, when dealing with a singular matrix, the presence of $k$ non-involutive constraints is anticipated. In our specific case, we have identified two involutive constraints, as indicated in \eqref{248}, which correspond to the null space of $M_{ij}$. Additionally, there are two non-involutive constraints detailed in \eqref{249}. Distinguishing between non-involutive and involutive constraints is crucial for understanding the system’s structure and behavior.
Furthermore, the new constraints derived from this approach have been validated using Dirac’s formalism \cite{Brown1}. This comparative analysis shows that first-class constraints are categorized as involutive, while second-class constraints are identified as non-involutive. This validation not only confirms the new constraints but also enhances our understanding of their role within the established theoretical framework.

Having fully identified the constraints, we now have the complete set given by $\left\{ H''_{1}, H''_{2}, H''_{3}, H''_{4} \right\}$. To integrate this set into the theoretical framework, we need to construct a differential that incorporates these constraints as generators of the system’s dynamics. However, the current parameter space does not fully encompass all constraints. Consequently, we will extend the parameter space by introducing new arbitrary variables $\left\{\omega, \tau, \lambda, \theta \right\}$. This extension enables us to define the system’s dynamics using the new fundamental differential

\begin{equation}\label{250}
	dF = \left\{F,H'_{0}\right\} dt + \left\{F,H''_{1}\right\}d\omega + \left\{F,H''_{2}\right\}d\tau + \left\{F,H''_{3}\right\} d\lambda + \left\{F,H''_{4}\right\} d\theta .
\end{equation}

The evolution of these new constraints gives
\begin{eqnarray}\label{251}
	dH''_{1} &=&H''_{2}dt= p_{3} dt = 0, \\
	dH''_{2}& =& 0, \\
	dH''_{3} &=& H''_{4}dt- d\theta=\left( \frac{p_{3}}{3} +q_{1} + q_{2} + q_{4} \right)dt  - d\theta = 0, \\
	dH''_{4} &=& p_{4}dt + d\lambda = 0.
\end{eqnarray}

%One can observe that $H''_{2}$ is obtained immediately by the evolution of $H''_{1}$. By the results obtained, we can note that the first two constraints are involutive, while the remaining two are not. 

%Due to the above, the GB is now introduced into the  fundamental differential to take the following form
It is evident that $H''_{2}$ is derived directly from the evolution of  $H''_{1}$. Based on the results obtained, we observe that the first two constraints are involutive, while the remaining two are not. Given this distinction, we now incorporate the GB into the fundamental differential. This integration modifies the differential as follows
\begin{equation}\label{252}
	dF = \left[ \left\{ F, H''_{\Bar{\alpha}} \right\} - \left\{ F, H''_{\Bar{a}} \right\} \left( M^{-1} \right)^{\Bar{a} \Bar{b}} \left\{ H''_{\Bar{b}}, H''_{\Bar{\alpha}} \right\}  \right]dt^{\Bar{\alpha}} =\left\{ F, H''_{\Bar{\alpha}} \right\}^{*} dt^{\Bar{\alpha}} ,
\end{equation}

here $\left( M^{-1} \right)^{\Bar{a} \Bar{b}}$  represents the inverse of the Poisson bracket matrix between non-involutive constraints, defined by $M_{\Bar{b} \Bar{a}} = \left\{ H''_{\Bar{b}}, H''_{\Bar{a}} \right\}$, with ${\Bar{a},\Bar{b}}=3,4$.  Additionally, $\Bar{\alpha} = 0,1,2$, with $t^{\Bar{0}} =t$, $t^{\Bar{1}} =\omega$, $t^{\Bar{2}} =\tau$, and we adopt the notation $H'_{0} \equiv H''_{0}$. This differential ensures that $dH''_{1} = dH''_{2} = dH''_{3} = dH''_{4}=0$. Consequently, we can use the second constraint from \eqref{248} to simplify the canonical Hamiltonian and proceed with the calculations in the GB framework. The resulting Hamiltonian will be
\begin{equation}\label{253}
	H_{0} = \frac{1}{2} \left[ p_{4}^{2} - \left(q_{1} + 2q_{2}\right)\left(q_{1} + 2q_{4}\right) \right] .
\end{equation}
%Now then, the characteristic equations obtained are the following
Now, the characteristic equations obtained are as follows
\begin{eqnarray}\label{254}
	dq_{1} &=& -\frac{1}{3} p_{4} dt + dt^{\Bar{1}}, \\
	dq_{2} &=& -\frac{1}{3} p_{4} dt -\frac{1}{2} dt^{\Bar{1}} , \\
	dq_{3} &=& \frac{1}{3} \left(p_{4} + q_{1} + q_{2} + q_{4}\right)dt + \frac{1}{2} dt^{\Bar{1}} + dt^{\Bar{2}},  \\
	dq_{4} &=& \frac{2}{3} p_{4} dt -\frac{1}{2} dt^{\Bar{1}},  \\
	dp_{1} &=& dp_{3} = 0, \\
	dp_{2} &=& (q_{4} - q_{2}) dt, \\
	dp_{4} &=& (q_{2} - q_{4}) dt.  \\
\end{eqnarray}

%Given that integrability is guaranteed, $t^{1},t^{2}$ and $t$ are linearly independent, leading to the following set of equations for time evolution

Given that integrability is assured, the variables $t^{\Bar{1}},t^{\Bar{2}}$, and $t$ are linearly independent. Consequently, the time evolution yields the following set of equations

\begin{eqnarray}\label{255}
	\dot{q}_{1} &=& - \frac{1}{3} p_{4}, \\
	\dot{q}_{2} &=& - \frac{1}{3} p_{4}, \\
	\dot{q}_{3} &=& \frac{1}{3} \left(p_{4} + q_{1} + q_{2} + q_{4}\right)= \frac{1}{3} p_{4} , \\
	\dot{q}_{4} &=& \frac{2}{3} p_{4}, \\
	\dot{p}_{2} &=& q_{4} - q_{2}, \\
	\dot{p}_{4} &=& q_{2} - q_{4}.
\end{eqnarray}

%We note that from the previous equations, the following relations are given: $\dot{q}_{4} = 2\dot{q}_{3}= -2\dot{q}_{2} = -2\dot{q}_{1}$ and $\dot{p}_{2} = -\dot{p}_{4}$, therefore, we can dispense with most of the degrees of freedom, retaining only two of them at the phase space level, since the solutions for the others can be immediately obtained through these relations.

From the previous equations, we observe the following relationships $\dot{q}_{4} = 2\dot{q}_{3}= -2\dot{q}_{2} = -2\dot{q}_{1}$ and $\dot{p}_{2} = -\dot{p}_{4}$.  Consequently, most of the degrees of freedom can be eliminated, retaining only two at the phase space level, as the solutions for the remaining variables can be directly derived from these relationships.

%Complementing this analysis, we also record the canonical transformations 

%The gauge transformations are given in the form of \eqref{23}, where the generating function for this case is

%In addition to this analysis, the corresponding canonical transformations are also derived according to $\delta F= \{F, H''_{z}\}dt^{z}$, for z=1,2,

In addition to this analysis, we also derive the corresponding canonical transformations using the relation $\delta F= \{F, H''_{z}\}dt^{z}$, for z=1,2,

\begin{eqnarray}\label{256}
    \delta q_{1} &=& \delta t^{\Bar{1}}, \quad 
	\delta q_{2} =-\frac{1}{2} \delta t^{\Bar{1}} , \quad
	\delta q_{3} =  \frac{1}{2} \delta t^{\Bar{1}} + \delta t^{\Bar{2}},  \quad 
	\delta q_{4} = -\frac{1}{2} \delta t^{\Bar{1}},  \quad  
	\delta p_{i} =  0. \nonumber \\
\end{eqnarray}

where the index  $i = 1,2,3,4$.

The point transformations given by equation \eqref{256} are determined by the generator function
\begin{equation}\label{257}
	G_{CF} = H''_{1} \delta t^{\Bar{1}} + H''_{2} \delta t^{\Bar{2}}.
\end{equation} 

%In this way, the infinitesimal gauge transformations are obtained:
%\begin{equation}
%	\delta q_{1} = \delta \phi,
%\end{equation}
%\begin{equation}
%	\delta q_{2} =	\delta q_{4} =-\frac{1}{2}\delta \phi,
%\end{equation}
%\begin{equation}
%	\delta q_{3} = \frac{1}{2}\delta \phi + \delta \psi,
%\end{equation}
%\begin{equation}
%	\delta q_{4} = -\frac{1}{2}\delta \phi,
%\end{equation}
%\begin{equation}
%	\delta p_{i} = 0,
%\end{equation}

%where the index  $i = 1,2,3,4$.

Next, we consider the first variation of the Lagrangian function \eqref{128} with respect to transformations of the form $\delta q_{i}\equiv q'_{i}-q_{i}$ is given by
%The lagrangian variation at constant time is:
\begin{equation}\label{258}
    \begin{split}
        \delta L = &\left( \dot{q}_{2} + \dot{q}_{3} + 2q_{1} + q_{2} + q_{4} \right)\delta q_{1} + \left( -2\ddot{q}_{2} - \ddot{q}_{3} + \ddot{q}_{4} - \dot{q}_{1} + q_{1} + 2q_{4} \right)\delta q_{2} \\
        &+ \left( -\ddot{q}_{2} - \ddot{q}_{3} - \dot{q}_{1} \right)\delta q_{3} + \left( \ddot{q}_{2} - \ddot{q}_{4} + q_{1} + 2q_{2} \right)\delta q_{4}\\
        &+ \frac{d}{dt} \left[ \left( q_{1} + 2\dot{q}_{2} + \dot{q}_{3} - q_{4} \right)\delta q_{2} + \left( q_{1} + \dot{q}_{2} + \dot{q}_{3} \right) \delta q_{3} + \left( \dot{q}_{4} - \dot{q}_{2} \right)\delta q_{4} \right].
    \end{split}
\end{equation}

%If \eqref{256} represent symmetries of the Lagrangian function, then $\delta L$ must equal zero. Substituting \eqref{256} into \eqref{270}, we have

%When evaluating the gauge transformations, one obtain that

If \eqref{256} represent symmetries of the Lagrangian function, then $\delta L$ must be zero. By substituting \eqref{256} into \eqref{258}, we obtain the following result
\begin{equation}\label{259}
\begin{split}
        \delta L &= \left( q_{1} + \dot{q}_{2} + \dot{q}_{3} \right)\delta \phi -\left( \dot{q}_{1} + \ddot{q}_{2} + \ddot{q}_{3} \right)\delta \psi + \frac{d}{dt} \left[\left( q_{1} + \dot{q}_{2} + \dot{q}_{3} \right)\delta \psi \right]\\
        & = \left( q_{1} + \dot{q}_{2} + \dot{q}_{3} \right)\left[ \delta \phi  + \frac{d}{dt} \left( \delta \psi \right) \right].
\end{split}
\end{equation}

%Hence, to comply that $\delta L = 0$, one must have: $\delta \phi = - d\left( \delta \psi \right)/dt $. Is is defined $\delta \psi \equiv \epsilon(t)$, so that $\delta \phi = - \dot{\epsilon}$. This implies that the generating function $G$ is then

Therefore, to ensure that $\delta L = 0$, it is necessary to have $\delta t^{\Bar{1}} = -  d\left( \delta t^{\Bar{2}} \right)/dt$. Let $\delta t^{\Bar{2}} = \epsilon(t)$, which implies $\delta t^{\Bar{1}} = - \dot{\epsilon}$. With these definitions, the generating function $G_{g}$ can be expressed as

\begin{equation}\label{260}
	G_{g} = -H''_{1} \dot{\epsilon} + H''_{2} \epsilon.
\end{equation} 

%This function $G_{g}$ serves as the generator of the gauge transformations given by

This function $G_{g}$ acts as the generator for the gauge transformations described by

\begin{eqnarray}\label{261}
	\delta q_{1} &=& -\dot{\epsilon}, \quad 
	\delta q_{2} =\frac{1}{2}\dot{\epsilon}, \quad
	\delta q_{3} = -\frac{1}{2}\dot{\epsilon} + \epsilon, \quad
    \delta q_{4} =\frac{1}{2}\dot{\epsilon}. 
\end{eqnarray}

It is important to emphasize that gauge transformations are derived by analyzing the relationship between canonical transformations and the symmetries of the Lagrangian. This approach does not employ Castellani's method, which is explored in \cite{Brown1} using Dirac's formalism. Nevertheless, despite the differing methodologies, both approaches lead to identical results.

Another important result from this formalism is the equivalence between Dirac brackets and GB, as well as the equations of motion in the reduced phase space. Given the inherent arbitrariness in the solutions of these equations, it becomes necessary to impose gauge conditions to resolve these ambiguities. This is achieved by introducing gauge conditions as constraints.

To illustrate this, consider the canonical gauge conditions specified by

\begin{eqnarray}\label{265}
H''_{5}&=&\chi_{1} = q_{1} - q_{2} = 0, \\
H''_{6}&=&\chi_{2} = q_{3} + p_{4} = 0.
\end{eqnarray}

In this context, applying these gauge conditions results in a set of non-involutive constraints. Consequently, the Poisson bracket matrix for these constraints will exhibit regularity. This matrix is denoted by  $M_{\mu \nu} = \left\{H''_{\mu}, H''_{\nu}\right\}$, where $\{\mu, \nu\} = {1, 2, 3, 4, 5, 6}$.

\begin{equation}\label{266}
		M_{\mu \nu} = 
		\begin{pmatrix}
			0 & 0 & 0 & 0 & -\frac{3}{2} & -\frac{1}{2}\\
			0 & 0 & 0 & 0 & 0 & -1 \\ 
			0 & 0 & 0 & -1 & 0 & \frac{1}{3} \\
			0 & 0 & 1 & 0 & 0 & \frac{2}{3} \\
			\frac{3}{2} & 0 & 0 & 0 & 0 & 0 \\
			 \frac{1}{2} & 1 & -\frac{1}{3} & -\frac{2}{3} & 0 & 0 
		\end{pmatrix} .
	\end{equation}

%Which is a non-singular matrix and calculating its inverse given as

We can remove the constraints and gauge conditions by formulating GB.
o achieve this, we use the inverse of the matrix $M_{\mu \nu}$ which is given by

\begin{equation}\label{266a}
M_{\mu \nu}^{-1} =
\begin{pmatrix}
0 & 0 & 0 & 0 & \frac{2}{3} & 0 \\
0 & 0 &  -\frac{2}{3} &  \frac{1}{3} &  -\frac{1}{3} & 1 \\
0 &  \frac{2}{3} & 0 & 1 & 0 & 0 \\
0 &  -\frac{1}{3} & -1 & 0 & 0 & 0 \\
-\frac{2}{3} &  \frac{1}{3} & 0 & 0 & 0 & 0 \\
0 & -1 & 0 & 0 & 0 & 0.
\end{pmatrix}
\end{equation}

Using this inverse matrix, the GB are defined by

\begin{equation}\label{266b}
	\left\{F, G\right\}^{**} =  \left\{ F, G \right\} - \left\{ F, H''_{\nu} \right\} \left( M^{-1} \right)^{\nu \mu} \left\{ H''_{\mu}, G\right\}. 
\end{equation}

%Now, we are going to find the evolution of the dynamic variables from the following differential:

%By doing so, the GB obtained will coincide with the Dirac brackets discussed in \cite{Brown1}

By applying this formula, the GB obtained will coincide with the Dirac brackets discussed in \cite{Brown1}. Specifically, the computed GB among the phase space variables are as follows

%By applying this formula, the GB obtained will coincide with the Dirac brackets discussed in \cite{Brown1}. Specifically, the BG  among the phase space variables are

\begin{equation}\label{268}
	\left\{q_{1},q_{3}\right\}^{**} =\left\{q_{1},p_{2}\right\}^{**} = -	\left\{q_{1},p_{4}\right\}^{**} = \frac{1}{3},
\end{equation}
\begin{equation}\label{269}
	\left\{q_{2},q_{3}\right\}^{**} =\left\{q_{2},p_{2}\right\}^{**} = -	\left\{q_{2},p_{4}\right\}^{**} = \frac{1}{3},
\end{equation}
\begin{equation}\label{270}
	\left\{q_{3},q_{4}\right\}^{**} =-\left\{q_{4},p_{2}\right\}^{**} = 	\left\{q_{4},p_{4}\right\}^{**} = \frac{2}{3}.
\end{equation}

These results demonstrate that the GBs align with the Dirac brackets as expected.

Now, we will examine the evolution of the dynamic variables using the following differential equation

\begin{equation}\label{267}
	dF = \left[ \left\{ F, H'_{0} \right\} - \left\{ F, H''_{\nu} \right\} \left( M^{-1} \right)^{\nu \mu} \left\{ H''_{\mu}, H'_{0}\right\} \right] dt = \left\{F,H'_{0}\right\}^{**}dt.
\end{equation}

%There, one can evaluate de gauge conditions on the hamiltonian $H'_{0}$ to reduce it, but also in the other constraints. Thus, the non-zero GB obtained are the same as those Dirac brackets from ref. \cite{Brown1}
In this context, gauge conditions can be applied to the Hamiltonian $H'_{0}$ helps simplify it, along with other constraints. Specifically, the Hamiltonian \eqref{133} simplifies to $H'_{0}=\frac{1}{2}(p^{2}_{2}+\frac{9}{4}q^{2}_{4})$. This simplified form coincides with the fully reduced Hamiltonian previously obtained in \cite{Brown1}.

The equations of motion obtained are
\begin{equation}\label{271}
	\dot{q}_{1} = \left\{q_{1},H'_{0}\right\}^{**} = -\frac{1}{3} p_{4},
\end{equation}
\begin{equation}\label{272}
	\dot{q}_{2} = \left\{q_{2},H'_{0}\right\}^{**} = -\frac{1}{3} p_{4},
\end{equation}
\begin{equation}\label{273}
	\dot{q}_{3} = \left\{q_{3},H'_{0}\right\}^{**} = q_{4} - q_{2},
\end{equation}
\begin{equation}\label{274}
	\dot{q}_{4} = \left\{q_{4},H'_{0}\right\}^{**} = \frac{2}{3} p_{4},
\end{equation}
\begin{equation}\label{275}
	\dot{p}_{1} = \left\{p_{1},H'_{0}\right\}^{**} = 0,
\end{equation}
\begin{equation}\label{276}
	\dot{p}_{2} = \left\{p_{1},H'_{0}\right\}^{**} = q_{4} - q_{2},
\end{equation}
\begin{equation}\label{277}
	\dot{p}_{3} = \left\{p_{3},H'_{0}\right\}^{**} = 0,
\end{equation}
\begin{equation}\label{278}
	\dot{p}_{4} = \left\{p_{4},H'_{0}\right\}^{**} =  q_{2} - q_{4}.
\end{equation}

%The relations between them are: $\dot{q}_{4} = - 2\dot{q}_{2}=- 2\dot{q}_{1}$ and $\dot{p}_{4} = -\dot{q}_{3} = -\dot{p}_{2}$. 

%so we could proceed exactly the same way as was done above to obtain solutions of $q_{4}$ y $p_{2}$.

From the set of equations \eqref{271}-\eqref{278}, we can establish relationships between them and obtain solutions in terms of  $q_{4}$ and $p_{2}$. These solutions correspond to a simple harmonic oscillator, characterized by the following equations

\begin{equation}
	q_{4}(t) = C_{1} \cos{t} -\frac{2}{3} C_{2} \sin{t} ,
\end{equation}
\begin{equation}
	p_{2}(t) = C_{2} \cos{t} +\frac{3}{2} C_{1} \sin{t}.
\end{equation}

To conclude this section, we will determine the degrees of freedom using two approaches. First, we will count the degrees of freedom by taking into account the constraints identified in the system, which include both involutive and non-involutive constraints. Second, we will count the degrees of freedom by considering gauge conditions as a set of additional constraints. This approach will illustrate how the introduction of gauge conditions results in a non-involutive system. It is crucial to note that, despite these differing methods, the total number of degrees of freedom remains consistent. The gauge conditions do not affect the overall count, ensuring that both methods yield the same number of degrees of freedom.

In the first approach, we have four constraints, of which two are non-involutive. The calculation of degrees of freedom is given by
\begin{equation*}
 \# \: \text{DOF} = \frac{1}{2} \left[ 8 - 2(2) - 2\right]  = 1.
\end{equation*}

In the second approach, by introducing the gauge conditions, we end up with six constraints, all of which are non-involutive. The corresponding calculation for the degrees of freedom is
\begin{equation*}
 \# \: \text{DOF} = \frac{1}{2} \left[ 8 - 2(0) - 6\right]  = 1.
\end{equation*}

Both methods yield the same number of degrees of freedom, confirming the consistency of the result across different approaches.

\section{Conclusions}

In this study, we conducted a comprehensive Hamilton–Jacobi analysis of classical dynamical systems with internal constraints to enhance our understanding of this formalism's application across various scenarios. Our investigation centered on four dynamic systems, specifically chosen for their relevance to the key principles under examination. Three of these systems included familiar components \cite{Brown2}, such as point masses connected by massless springs, rods, ropes, and pulleys. By selecting these well-known setups, we were able to apply the Hamilton-Jacobi formalism to established scenarios, ensuring that our analysis remained clear and accessible.

Transitioning to a more complex scenario, we selected the fourth system, previously examined by David Brown \cite{Brown1}, due to its effectiveness in illustrating the full logical framework of the Dirac-Bergmann algorithm. This choice was essential, as it allowed for a direct comparison with the Hamilton-Jacobi approach used in our study. We organized our analysis methodically, focusing on the constraints present in each system: the first two systems featured non-involutive constraints, the third system was characterized by involutive constraints, and the fourth system presented a combination of both types. This organization allowed us to systematically determine the degrees of freedom for each system and derive the corresponding equations of motion, thus laying the groundwork for our subsequent analysis.

To address and eliminate non-involutive constraints effectively, we employed generalized brackets, which played a key role in ensuring the integrability of the theory. The integrability condition in Dirac's formalism is equivalent to its consistency conditions. Specifically, satisfying these consistency conditions implies that \( dH'_{\alpha} = 0 \) \cite{Pi2, Bravo, Heredia}. This establishes a direct link between integrability and the formalism's consistency. This approach enabled us to systematically manage complex constraint structures, facilitating the formulation of accurate equations of motion. In sections III and IV, where we focused on systems with only non-involutive constraints, the challenge of finding an analytical solution to the Hamilton-Jacobi Partial Differential Equations (HJPDEs) was evident. Therefore, we relied on the equations of motion derived from the Generalized Bracket (GB) approach. This reliance underscored the equivalence between the Dirac method and the Hamilton-Jacobi formalism in handling such constraints, as both methodologies prioritize the Canonical Hamiltonian and its potential reductions without resorting to the construction of a Total Hamiltonian.

Building on the foundation established in the previous sections, section V shifted the focus to a completely involutive system with a single constraint derived from the null vector associated with the Hessian matrix. This analysis set the stage for section VI, where a more intricate challenge arose: the reclassification of initially obtained constraints due to their linear dependence. This reclassification led to the identification of two involutive and two non-involutive constraints. For the systems discussed in sections V and VI, which involved involutive constraints, we delved into the relationship between characteristic flows and Lagrangian point (gauge) transformations. This deeper exploration was instrumental in identifying the gauge generator and the corresponding gauge transformations for these systems, as detailed in \eqref{126} and \eqref{261}.

These varied examples highlight the broad spectrum of possible constraints, encompassing involutive, non-involutive, and mixed cases. Notably, in the latter example, we employed two distinct methods to calculate the equations of motion and perform degree of freedom counting. Initially, a direct analysis of the constraints revealed two involutive and two non-involutive constraints with one degree of freedom. Subsequently, after implementing gauge fixing, all constraints became non-involutive, resulting in six non-involutive constraints while maintaining the same number of degrees of freedom. By comparing these results with those reported in \cite{Brown1, Brown2,Arellano}, we ensured consistency with prior studies, reinforcing the reliability of our findings.

To validate the robustness of our conclusions, we compared the results obtained through the Hamilton-Jacobi approach with those derived from the Dirac-Bergmann \cite{Brown1,Brown2} and Faddeev-Jackiw algorithms \cite{Arellano}. This comparative analysis confirmed the effectiveness of the Hamilton-Jacobi method, demonstrating its simplicity and efficiency in mathematical operations. Furthermore, the method's ease of implementation on computational platforms highlights its practicality, making it a valuable tool for applications in computational physics.

In summary, our study reinforces the Hamilton-Jacobi approach as a robust and adaptable framework for analyzing classical dynamical systems with internal constraints, providing both theoretical insights and practical benefits that extend across various domains of physics.

\newpage
\section*{APPENDIX: CODE IMPLEMENTATION IN WOLFRAM MATHEMATICA FOR THE CALCULATION OF RESULTS}

The Hamilton-Jacobi formalism extended for hamiltonian systems with internal constraints has the virtue of containing expressions within its framework that permit devising a way to implement it into a symbolic computation language. In this context, the classification of constraints becomes the main and necessary task to subsequently obtain the expected final results, such as the Characteristic Equations and the counting of degrees of freedom. To implement the Hamilton-Jacobi formalism in a symbolic computation language like Mathematica, a set of functions that allow replicating the results obtained manually have been developed.

In the following pages, pseudocodes, codes, and explanations related to these functions developed in the Wolfram Mathematica symbolic computation language will be presented. These functions allowed us to replicate the relevant results that manual calculations gave us, which coincide with the corresponding ones presented in the literature.

Before delving into the details of these functions, it is essential to understand how Mathematica functions operate. It is well known that Mathematica comes with several built-in functions that provide different facilities. However, the Wolfram Language allows the creation of new functions based on the existing ones. To evaluate every function in Mathematica, one must specify its arguments; knowing the correct way to declare and introduce the arguments for the functions about to be shown is of utmost importance. The following specifications are crucial for ensuring that the functions perform correctly:

\begin{itemize} 

\item The list of dynamic variables must always be ordered: first the $N$ coordinates, and then the $N$ conjugate momentum in the same respective order. In the end, there will be $2N$ elements, such that the coordinate in the $j$-th position has its conjugate momentum in the $(j+N)$-th position. In this appendix, the list could be found labeled as \textit{variables}.

\item The \textit{constraints} list's elements must be in ascending order. 

Example: $\textit{constraints=}$$\left\{H1,H2,H3,...\right\}$.

\item The list labeled as \textit{HJPDES} is almost the same as the constraints list, with the difference that the term $H0 = p0+h0$ goes first, where $h0$ is the label for the canonical Hamiltonian. One could also just write the canonical Hamiltonian and omit $p0$. It is just written to remind that in the HJ formalism the notation associated is: $p0 = \partial S/\partial t$, although this term has no influence on the calculations.

\item Differentials of independent variables must also be inserted in a list in the same order as their associated Hamiltonians appear in the \textit{HJPDES} list. Example: independent variables associated with the Hamiltonians: $H0 \rightarrow t$ , $H1 \rightarrow q1$, $H2 \rightarrow q4$, $H3 \rightarrow \omega$, $H4 \rightarrow \psi$, so that: the \textit{HJPDES} list is: $\left\{H0,H1,H2,H3,H4\right\}$ and the indvar list of differentials of independent variables is $\left\{dt,dq1,dq2,d\omega, d\psi\right\}$.

\end{itemize}

\newpage

After establishing the importance of properly configuring the lists, we now turn our attention to the specific functions responsible for performing key calculations within the theory. These functions are designed to calculate various mathematical objects derived from the theory:

\begin{itemize}
    \item The Poisson bracket
%\begin{figure}[h]
 %   \centering
  %  \includegraphics[scale=1]{Mathematica functions prints/PoissonBracket.pdf}
    %\caption{Caption}
    %\label{fig:enter-label}
%\end{figure}

Calculates the Poisson Bracket of between two variables/functions \textit{f} and \textit{g}.

\item The fundamental differential
%\begin{figure}[h]
%    \centering
%    \includegraphics[scale=1]{Mathematica functions prints/Differential.pdf}
    %\caption{Caption}
    %\label{fig:enter-label}
%\end{figure}

Calculates the fundamental differential of a given variable \textit{f}, in the form of eq. \eqref{13}, and making the Poisson brackets with respect the list of hamiltonians labeled as \textit{hjpdes}. It incorporates its corresponding independent variables from the list \textit{indvar}.

\item The Poisson bracket matrix between constraints
%\begin{figure}[h]
 %   \centering
  %  \includegraphics[scale=1]{Mathematica functions prints/PBMatrix.pdf}
    %\caption{Caption}
    %\label{fig:enter-label}
%\end{figure}

Makes all the possible combinations of Poisson brackets between a set of constraints and arrange the results in a matrix input form.

%\newpage

    \item The Generalized Bracket

%\begin{figure}[h]
%    \centering
%    \includegraphics[scale=1]{Mathematica functions prints/GeneralizedBracket.pdf}
    %\caption{Caption}
    %\label{fig:enter-label}
%\end{figure}

Calculates the GB between two variables/functions \textit{f} and \textit{g}, It has to take the given list \textit{constraints} in its argument [See eq. \eqref{18}].

    \item The fundamental differential for partially involutive systems

%\begin{figure}[h]
 %   \centering
  %  \includegraphics[scale=1]{Mathematica functions prints/DiffPartiallyInvolutive.pdf}
    %\caption{Caption}
    %\label{fig:enter-label}
%\end{figure}

To accurately calculate the fundamental differential of a given variable or function \textit{f} [see eq. \eqref{20}], it is essential to properly organize the differentials of the independent variables within the argument \textit{indvarsreducedlist}. This list must include all differentials corresponding to the involutive Hamiltonians from the \textit{invconstraints} list, ensuring that $H0$ is the first element and that the time differential $dt$ is placed at the beginning of the \textit{indvarsreducedlist}.
\end{itemize}

%\newpage

Moreover, in addition to correctly organizing the differentials, it is equally important to present these results in a clear and readable format. The \texttt{SuscriptableSymbols} and \texttt{MechSub} functions were developed to print results in subscript format, drawing inspiration from the early modules of the \textbf{SymbolToSubscript} resource function (see ref. \cite{Chan}). %These functions are defined as follows
%\begin{figure}[h]
   % \centering
    %\includegraphics[scale=1]{Mathematica functions prints/SuscriptableSymbols.pdf}
    %\caption{Caption}
    %\label{fig:enter-label}
%\end{figure}
%\vspace{-0.7cm}
%\begin{figure}[h]
%    \centering
%    \includegraphics[scale=1]{Mathematica functions prints/MechSub_2.pdf}
    %\caption{Caption}
    %\label{fig:enter-label}
%\end{figure}

%\newpage
%\begin{figure}[h]
%    \centering
%    \includegraphics[scale=1]{Mathematica functions prints/MechSub_3.pdf}
    %\caption{Caption}
    %\label{fig:enter-label}
%\end{figure}
After introducing the \texttt{MechSub} function, it is important to discuss additional functions that further enhance the organization of results. In particular, the functions \texttt{NegGroupHold}, \texttt{PrintTraditionalOrder}, and \texttt{OrganizeTerms} collaborate to present results with factored terms in a format that is more structured and organized compared to the software’s default output. The definitions of these functions are as follows

%\begin{figure}[h]
 %   \centering
  %  \includegraphics[scale=1]{Mathematica functions prints/TradOrd_1.pdf}
    %\caption{Caption}
    %\label{fig:enter-label}
%\end{figure}

%\begin{figure}[h]
%    \centering
 %   \includegraphics[scale=1]{Mathematica functions prints/TradOrd_2.pdf}
    %\caption{Caption}
    %\label{fig:enter-label}
%\end{figure}

\newpage
%\begin{figure}[!h]
 %   \centering
 %   \includegraphics[scale=1]{Mathematica functions prints/TradOrd_3.pdf}
    %\caption{Caption}
    %\label{fig:enter-label}
%\end{figure}

The \texttt{MechSub} function, along with the three previously mentioned functions, does not significantly affect the calculations. Their sole purpose is to enhance the aesthetics and organization of the printed results.

The core of this computational approach is encapsulated in a modular function called \texttt{SystemInvolutionAnalysis}. To utilize this function, the user must provide the canonical Hamiltonian and at least a set of primary constraints, which need to be obtained beforehand.

The first step when using this function is for the user to specify whether the constraints list provided is the final set of constraints for the system or merely the primary constraints. If the former is true, the user should respond with \textit{Yes}. If the latter is true, or if the user is unsure, they should respond with \textit{No} or \textit{I don't know}. In this case, the function will only generate secondary constraints and treat the primary and secondary constraints as the system's final set. It’s crucial to note that the function is only capable of automatically generating secondary constraints. If errors occur during its operation, it may indicate that the system has additional constraints that need to be manually derived using the initial functions outlined in this appendix and the theoretical framework provided.

Once a final set of constraints is established, these should be included in the \textit{constraints} and \textit{hjpdes} lists, along with the corresponding differentials of independent variables in the \textit{indvar} list. Another key point is that the function does not automatically simplify the results by evaluating the constraints, so the user may need to manually evaluate the necessary constraints to achieve the most simplified expressions. 

Following this, the structure of the \texttt{SystemInvolutionAnalysis} function is detailed below. To enhance understanding, we provide pseudocode that outlines the algorithmic steps involved in constructing the \texttt{SystemInvolutionAnalysis} function. This pseudocode offers a concise representation of the underlying mathematical framework.
%The structure of the \texttt{SystemInvolutionAnalysis} function is as follows
%%%%%%%%%%%%%%%%% P S E U D O C O D E %%%%%%%%%%%%%%%%%%

\begin{algorithm*}
%\caption{Analysis of the system attributes and calculus of CE's}
\caption{SystemInvolutionAnalysis}
\begin{algorithmic}[1]
    \Function{SystemInvolutionAnalysis}{$constraints$, $hjpdes$, $indvar$, $variables$}

    \State Variable \textit{answer}, \textit{constraints1}, \textit{hjpdes1}, \textit{indvars}, \textit{noninvconstraints}, 
\textit{invconstraints}, \textit{indvarreducedlist}
    
        \If{$answer = $ \textit{"No"} $||$ $answer =$ \textit{"I don't know"}}        

           ($constraints1$, $hjpdes1$,$indvars$) $\gets$ CheckNewConstraints($constraints$, $hjpdes$, $indvar$, $variables$)
               
        \Else 
        
            ($constraints1$, $hjpdes1$,$indvars$) $\gets$ ($constraints$, $hjpdes$,$indvar$) 
            
        \EndIf
   
        ApplyDifferentialToConstraints($constraints1$, $hjpdes1$, $indvars$, $variables$)

       ($invconstraints$, $noninvconstraints$, $indvarreducedlist$) $\gets$ ClassifyConstraints($constraints1$, $hjpdes1$, $indvars$, $variables$)

        ApplyOperations($invconstraints$, $noninvconstraints$,$indvarreducedlist$,$hjpdes1$, $indvars$, $variables$)

        CountDegreesOfFreedom($variables$, $constraints1$, $noninvconstraints$)

        PrintFinalConstraints($constraints1$)

        SeparateConstraints($invconstraints$, $noninvconstraints$)

    \EndFunction
\end{algorithmic}
\end{algorithm*}

%%%%%%%%%%%%%%%%%%%%%%%%%%%%%%%%%%%%%%%%%%%%%%%%%%%%%%

\newpage

%The code structure, written in Mathematica, is given by

%\begin{figure}[h]
%    \centering
%    \includegraphics[scale=1]{Mathematica functions prints/SystemInvolutionAnalysisFUNCTION.pdf}
    %\caption{Caption}
    %\label{fig:enter-label}
%\end{figure}

%Each module follows the coding structure outlined below. The modules \texttt{ApplyDifferentialToConstraints} and \texttt{ApplyOperations} contain in its structures some modifications of the resource function \textbf{SubscriptedSymbols} (See ref. \cite{Chan1}).
Each module adheres to the coding structure outlined below. 

The modules \texttt{ApplyDifferentialToConstraints} and \texttt{ApplyOperations} incorporate modifications of the \textbf{SubscriptedSymbols} resource function (see ref. \cite{Chan2}).

The \texttt{SystemInvolutionAnalysis} modules are addressed:

\begin{itemize}
\item \texttt{CheckNewConstrains}
\end{itemize}

The function operates by first checking if the results from applying \texttt{Differential} to all primary constraints yield only one non-zero Poisson bracket. If so, this Poisson bracket is identified as a secondary constraint and added to a new local list named \textit{constraints1}, which will include both primary and secondary constraints. Additionally, any new Hamiltonians are added to another local list, \textit{hjpdes1}, which should have corresponding new independent variables. Consequently, the differentials of these variables are included in the local list \textit{indvars}. Finally, the function outputs an updated local list containing these three lists: \textit{constraints1}, \textit{hjpdes1}, and \textit{indvars}.

%\newpage

%%%%%%%%%%%%%%%%% P S E U D O C O D E %%%%%%%%%%%%%%%%%%
\begin{algorithm*}
\caption{CheckNewConstraints}
\begin{algorithmic}[1]
\Function{CheckNewConstraints}{$constraints$, $hjpdes$, $indvar$, $variables$}
    \State Variable \textit{result}, \textit{newconstr}, \textit{constraints1} $\gets$ \textit{constraints},  \textit{hjpdes1} $\gets$\textit{hjpdes}, \textit{indvars} $\gets$ \textit{indvar}, \textit{gcharacters} $\gets$ \{Greek characters from $\alpha$ to $\omega$ excluding $\pi$\}
    %\State \textit{newconstr}
    %\State \textit{constraints1} $\gets$ \textit{constraints}
    %\State \textit{hjpdes1} $\gets$\textit{ hjpdes}
    %\State \textit{indvars} $\gets$ \textit{indvar}
    %\State \textit{gcharacters} $\gets$ \{Greek characters from $\alpha$ to $\omega$ excluding $\pi$\}
    \For{$i \gets 1$ to length of \textit{constraints}}
        \State \textit{result} $\gets$  %\Function{Differential}{$constraints[i]$, $hjpdes$, $indvar$, $variables$}
        %\EndFunction
         $\sum_{\alpha} \left\{ q_{i}, H'_{\alpha} \right\}dt^{\alpha}$
        \If {\textit{result} matches pattern \textit{newconstr} * \textit{elem}, where \textit{elem} is in \textit{indvar}}
            \State \textit{newconstr} $\gets$ Replace \textit{elem} with 1 in \textit{result}
            \State Append \textit{newconstr} to \textit{constraints1}
            \State Append \textit{newconstr} to \textit{hjpdes1}
            \State Append a new symbol "d" followed by a random Greek character from \textit{gcharacters} to \textit{indvars}
        \EndIf
    \EndFor
    \State \Return $\{constraints1, hjpdes1, indvars\}$
\EndFunction
\end{algorithmic}
\end{algorithm*}
%%%%%%%%%%%%%%%%%%%%%%%%%%%%%%%%%%%%%%%%%%%%%%%%%%%%%%

%In Mathematica this was coded :

%\begin{figure}[h]
%    \centering
 %   \includegraphics[scale=0.9]{Mathematica functions prints/SysInvAnModule2_CheckNewConstraints.pdf}
    %\caption{Caption}
    %\label{fig:enter-label}
%\end{figure}

%The present function ensures that if the results obtained from applying \texttt{Differential} to all primary constraints have just one non-zero Poisson bracket, then the result from that Poisson bracket gets recognized as a secondary constraint and added to new local list of constraints named \textit{constraints1} that will include primary and secondary. Also, those new hamiltonians will be included into the local list \textit{hjpdes1}, which should have associated new independent variables, and therefore, its differentials into the local list \textit{indvars}. In the end, the output of this function is a local list of these three updated lists.

 %\newpage

\begin{itemize}
\item \texttt{ApplyDifferentialToConstraints}
\end{itemize}

In brief, this function applies \texttt{Differential} to a final set of constraints, including both the primary constraints and any additional constraints derived from them.

After processing these constraints, the function's main task is to present the results in a labeled format, specifically using subindex notation. This ensures that the output is clearly organized and easily interpretable.

\newpage

\begin{algorithm*}
    \caption{ApplyDifferentialToConstraints}
    \begin{algorithmic}[1]
    \Function{ApplyDifferentialToConstraints}{$constraints$, $hjpdes$, $indvars$, $variables$}

    \State Variable \textit{label} $\gets \{ \}$, \textit{result} $\gets \{ \}$
    
        \For{$i \gets$ 1 to  Length of \textit{constraints}}

        Append $dH_{i}$ to \textit{label}
        
        Append $\sum_{\alpha} \left\{ H'_{i}, H'_{\alpha} \right\}dt^{\alpha}$ to \textit{result}
        \EndFor

        \For{$i \gets$ 1 to Length of \textit{constraints}}
        
        \State Print "\textit{label}[i] == \textit{result}[i]"
        
        \EndFor

    \EndFunction
    \end{algorithmic}
\end{algorithm*}

%This is done in the Wolfram Language as showed:

%\begin{figure}[!h]
%    \centering
%    \includegraphics[scale=0.9]{Mathematica functions prints/SysInvAnModule3_ApplyDifferentialToConstraints.pdf}
    %\caption{Caption}
    %\label{fig:enter-label}
%\end{figure}

%In brief, what this functions does is applying \texttt{Differential} to a final set of constraints. i.e., to the primary constraints and all constraints able to obtain subsequent to the primary ones.

%The main task of this function is printing these results in a labeled form with the sub index format.

%\newpage

\begin{itemize}
\item \texttt{ClassifyConstraints}
\end{itemize}

When analyzing the differential of the final set of constraints, this module generates two distinct local lists to categorize the constraints: involutive and non-involutive. These lists are named \textit{invconstraints} and \textit{noninvconstraints}, respectively.

The module assesses whether the result of applying \texttt{Differential} is zero or non-zero. If the result is zero, it further examines whether there might be non-zero Poisson brackets indicating constraints within the same final set. Such constraints should be correctly categorized as zero if they are identified. 

In addition, the function includes the differentials of independent variables associated with involutive Hamiltonians in a local list named \textit{indvarreducedlist}. 

The final output of this module is a comprehensive list comprising the three aforementioned lists.

%%%%%%%%%%%%%%%%% P S E U D O C O D E %%%%%%%%%%%%%%%%%%
\begin{algorithm*}
\caption{ClassifyConstraints}
\begin{algorithmic}[1]
\Function{ClassifyConstraints}{$constraints$, $hjpdes$, $indvars$, $variables$}
    \State Variable \textit{invconstraints} $\gets$ \{\textit{hjpdes}[1]\},  \textit{noninvconstraints} $\gets$ \{\}, \textit{indvarreducedlist} $\gets$ \{$dt$\}
    %\State \textit{indvarreducedlist} $\gets$ \{$dt$\}
    %\State \textit{noninvconstraints} $\gets$ \{\}
    %\State \textit{indvarreducedlist} $\gets$ \{$dt$\}
    \For{$i \gets 1$ to Length of \textit{constraints}}
        \If{
            %\Function{Differential}{($constraints$[i], $hjpdes$, $indvars$, $variables$)}
            %\EndFunction
            $\sum_{\alpha} \left\{ q_{i}, H'_{\alpha} \right\}dt^{\alpha}$ matches a member from a list of results of differentials in which all the non-zero Poisson brackets are equal to constraints.
            
            %Map(Total, Subsets(Flatten(Outer(\#*\#2\&,constraints, indvars)))) Or Map(Total, Subsets(Flatten(Outer((-\#)*\#2\&,constraints, indvars))))
        } 
    
        Append \textit{constraints}[i] to \textit{invconstraints}

        Append \textit{indvars}[$i + 1$] to \textit{indvarreducedlist}
    \Else

        Append \textit{constraints} [i] to \textit{noninvconstraints}
        
    \EndIf
\EndFor
    \State \Return \{\textit{invconstraints}, \textit{noninvconstraints}, \textit{indvarreducedlist}\}

\EndFunction
\end{algorithmic}
\end{algorithm*}
%%%%%%%%%%%%%%%%%%%%%%%%%%%%%%%%%%%%%%%%%%%%%%%%%%%%%%%

%Wolfram Language:

%\begin{figure}[h]
 %   \centering
 %   \includegraphics[scale=1]{Mathematica functions prints/SysInvAnModule4_ClassifyConstraints1.pdf}
    %\caption{Caption}
    %\label{fig:enter-label}
%\end{figure}

%\begin{figure}[h]
%    \centering
%    \includegraphics[scale=1]{Mathematica functions prints/SysInvAnModule4_ClassifyConstraints2.pdf}
    %\caption{Caption}
    %\label{fig:enter-label}
%\end{figure}

%When analyzing the differential of the final set of constraints, this module will generate two distinct local lists to classify the constraints into the two types: involutive and non-involutive, naming those lists \textit{invconstraints} and \textit{noninvconstraints}, respectively. The judgment made over the constraints consists on revising if the result of applying \texttt{Differential} is different or equal to zero (in this last case, taking in account if there could be non-zero Poisson bracket that result in being constraints form the same final set, which evidently are set to be zero). This function will also add the respective differentials of independent variables associated with the involutive hamiltonians to the local list named \textit{indvarreducedlist}. The output is a local list of the three lists said before.

\newpage

\begin{itemize}
\item \texttt{ApplyOperations}
\end{itemize}

This module calculates the Characteristic Equations for all dynamic variables in the complete phase space, tailored to the system's involution. It does this by examining the number of elements in the two constraint lists produced by the previous function, \texttt{ClassisfyConstraints}. If the list \textit{noninvconstraints} has zero elements, it indicates that the system is fully involutive. In this case, the Characteristic Equations are computed using \texttt{Differential}.

In contrast, if the length of \textit{noninvconstraints} equals the total number of Hamiltonians minus one (the canonical Hamiltonian), it suggests that the system is non-involutive. For such cases, the Characteristic Equations are determined using \texttt{GeneralizedBracket}, with time $t$ as the only independent parameter.

If neither of these conditions is met, the system is classified as partially involutive. Here, the Characteristic Equations are calculated using \texttt{DiffPartiallyInvolutive}, which relies on a set of independent variables linked to the involutive Hamiltonians.

Finally, this module outputs the Characteristic Equations in the specified subindex format. The function \texttt{OverOneDotFunc} is included to facilitate this process.

%%%%%%%%%%%%%%%%% P S E U D O C O D E %%%%%%%%%%%%%%%%%%
\begin{algorithm*}
    \caption{ApplyOperations}
    \begin{algorithmic}[1]
    \Function{ApplyOperations}{$invconstraints$, $noninvconstraints$, $indvarreducedlist$, $hjpdes$, $indvars$, $variables$}
            \State Variable eqnsci $\gets$ \{ \}, labsci $\gets$  \{ \}, eqnsni $\gets$  \{ \}, labsni $\gets$  \{ \}, eqnspi $\gets$  \{ \}, labspi $\gets$  \{ \}
        \If{Length of \textit{noninvconstraints} equals 0}

           \State Print "The system is completely involutive. The Characteristic Equations are obtained through the Fundamental Differential, therefore the results are: "
           \For{i $\gets$ 1 to Length of \textit{variables}}
           Append $dq_{i}$ to \textit{labsci} 
           \EndFor

            \For{i $\gets$ 1 to Length of \textit{variables}}
            Append $ \sum_{\alpha} \left\{ q_{i}, H'_{\alpha} \right\}dt^{\alpha}$ to        
            \textit{eqnsci}
            \EndFor

            \For{i $\gets$ 1 to Length of \textit{variables}}
            Print "\textit{labsci}[i] == \textit{eqnsci}[i]"
            \EndFor

            \Else
                \If{Length of \textit{noninvconstraints} equals (Length of \textit{invconstraints} + Length of \textit{noninvconstraints} - 1) }

                \State Print "The system is non-involutive. The Characteristic Equations are obtained through the Generalized Bracket, therefore the results are: "

                \For{i $\gets$ 1 to Length of \textit{variables}}
                Append $\dot{q}_{i}$ to \textit{labsni} 
                \EndFor

                \For{i $\gets$ 1 to Length of \textit{variables}}
                Append $\left\{ q_{i}, H'_{0} \right\}^{*} dt$ to \textit{eqnsni} 
                \EndFor

                \For{i $\gets$ 1 to Length of \textit{variables}}
                Print "\textit{labsni}[i] == \textit{eqnsni}[i]"
                \EndFor
                
                \Else
                    \State  Print "The system is partially involutive. The Characteristic Equations are obtained through Generalized Brackets, therefore the results are: 

                    \For{i $\gets$ 1 to Length of \textit{variables}}
                    Append $\dot{q}_{i}$ to \textit{labspi} 
                    \EndFor

                    \For{i $\gets$ 1 to Length of \textit{variables}}
                    Append $ \sum_{\Bar{\alpha}} \left\{ q_{i}, H'_{\Bar{\alpha}} \right\}^{*}dt^{\Bar{\alpha}}$ to \textit{eqnspi}
                    \EndFor

                    \For{i $\gets$ 1 to Length of \textit{variables}}
                    Print "\textit{labspi}[i] == \textit{eqnspi}[i]"
                    \EndFor

                \EndIf

        \EndIf

    \EndFunction   
    \end{algorithmic}
\end{algorithm*}
%%%%%%%%%%%%%%%%%%%%%%%%%%%%%%%%%%%%%%%%%%%%%%%%%%%%%%

\newpage

\newpage
\begin{itemize}
\item \texttt{OverOneDotFunc} 
\end{itemize} 

%\begin{figure}[!h]
 %   \centering
  %  \includegraphics[scale=0.9]{Mathematica functions prints/SysInvAnModule1_OverOneDotFunc.pdf}
    %\caption{Caption}
    %\label{fig:enter-label}
%\end{figure}

This brief function permits printing results with the Newton notation for derivatives.

\begin{itemize}
\item \texttt{CountDegreesOfFreedom}
\end{itemize}

This module is responsible for calculating and printing the number of degrees of freedom, as defined by equation \eqref{DOF}. 
%It is specified in the following form

%\begin{figure}[h]
%    \centering
%    \includegraphics[scale=0.9]{Mathematica functions prints/SysInvAnModule6_CountDegreeOfFreedom.pdf}
    %\caption{Caption}
    %\label{fig:enter-label}
%\end{figure}

%This module is in charge of calculating and printing the number of degrees of freedom as given by eq. \eqref{DOF}.

\begin{itemize}
\item \texttt{PrintFinalConstraints}
\end{itemize}
%\begin{figure}[!h]
 %   \centering
  %  \includegraphics[scale=1]{Mathematica functions prints/SysInvAnModule7_PrintFinalConstraints.pdf}
    %\caption{Caption}
    %\label{fig:enter-label}
%\end{figure}

A list of the final set of constraints is printed by this module.

%\newpage

\begin{itemize}
\item \texttt{SeparateConstraints}
\end{itemize}

The final module outputs a message that classifies all constraints, distinguishing between involutive constraints ("first-class") and non-involutive constraints ("second-class").

\begin{algorithm*}
    \caption{SeparateConstraints}
    \begin{algorithmic}[1]
    \Function{SeparateConstraints}{$invconstraints$, $noninvconstraints$}
        \If{Length of \textit{noninvconstraints} equals 0}
        Print "All constraints are 'first-class' "
        \Else
            \If{Length of \textit{invconstraints} equals 1}
            Print "All constraints are  'second-class'"
            \Else
    
               \State Print "The 'first-class' constraints are:  
                *\textit{invconstraints} with its first element deleted*,
                while the 'second-class' constraints are: \textit{noninvconstraints}"

            \EndIf
                
        \EndIf

    \EndFunction
    \end{algorithmic}
\end{algorithm*}

%Wolfram Language:
%\begin{figure}[h]
 %   \centering
  %  \includegraphics[scale=1]{Mathematica functions prints/SysInvAnModule8_SeparateConstraints.pdf}
    %\caption{Caption}
    %\label{fig:enter-label}
%\end{figure}

%This last module prints a message indicating the classification given of all constraints, establishing which are the involutive ones ("first-class") and the non-involutive ones ("second-class").

%\newpage

%An example of its execution is the one that will be presented below. It corresponds to the system from section IV: Masses, springs and rings.

An example of this procedure will be presented below, corresponding to the system described in Section IV: Masses, Springs, and Rings.

First, it is necessary to declare the variables required for the \texttt{SystemInvolutionAnalysis} function. The list of dynamic variables is defined as $variables = \left\{ x,y,\theta_1, \theta_2, \theta_3, p_x, p_y, p_1, p_2, p_3 \right\}$. Next, the canonical Hamiltonian $H_0$ [Eq. \eqref{78}] labeled as $h0$, along with its corresponding canonical constrain $H'_0$ [Eq. \eqref{78a}], labeled as $H0$. Following this, the primary constraints $H'_1$ [Eq. \eqref{78b}] and $H'_2$ [Eq. \eqref{78c}], labeled as $H1$ and $H2$ respectively, are introduced, with the associated independent variables being $x$ and $y$. Finally, the list of Hamiltonians is defined as $HJPDES =\left\{H0, H1, H2 \right\} $, accompanied by the corresponding list of differentials of independent variables $indvar = \left\{dt, dx, dy \right\}$, and the list of primary constraints $constraints = \left\{H1, H2 \right\}$. In the Mathematica environment, these declarations are executed as follows

\begin{figure}[h]
    \centering
    \includegraphics[scale=1]{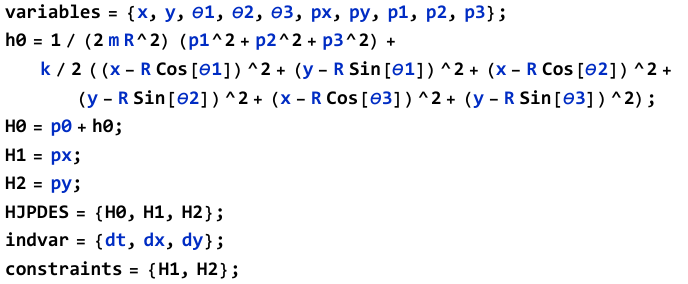}
    %\caption{Caption}
    %\label{fig:enter-label}
\end{figure}

Once the necessary variables and constraints have been defined, the function is executed with the corresponding arguments

\begin{figure}[h]
    \centering
    \includegraphics[scale=1]{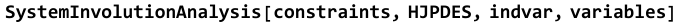}
    %\caption{Caption}
    %\label{fig:enter-label}
\end{figure}

The results obtained from the execution are as follows

\begin{figure}[!h]
    \centering
    \includegraphics[scale=1]{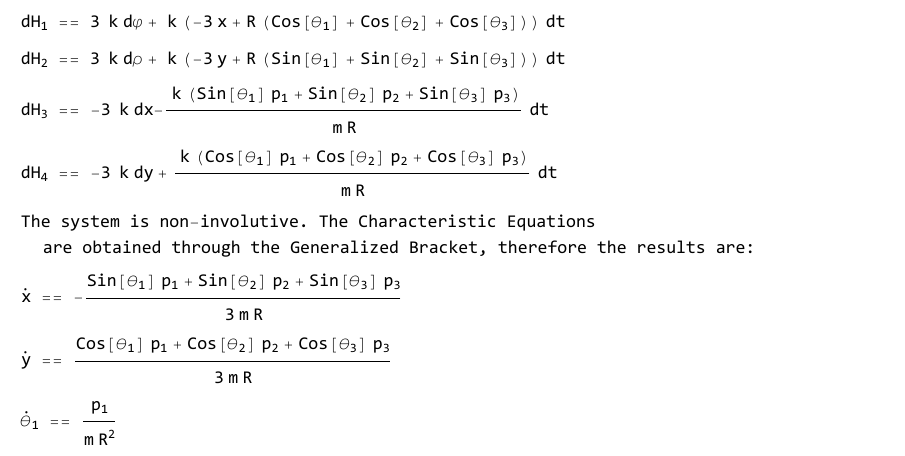}
    %\caption{Caption}
    %\label{fig:enter-label}
\end{figure}

%\newpage

\begin{figure}[!h]
    \centering
    \includegraphics[scale=1]{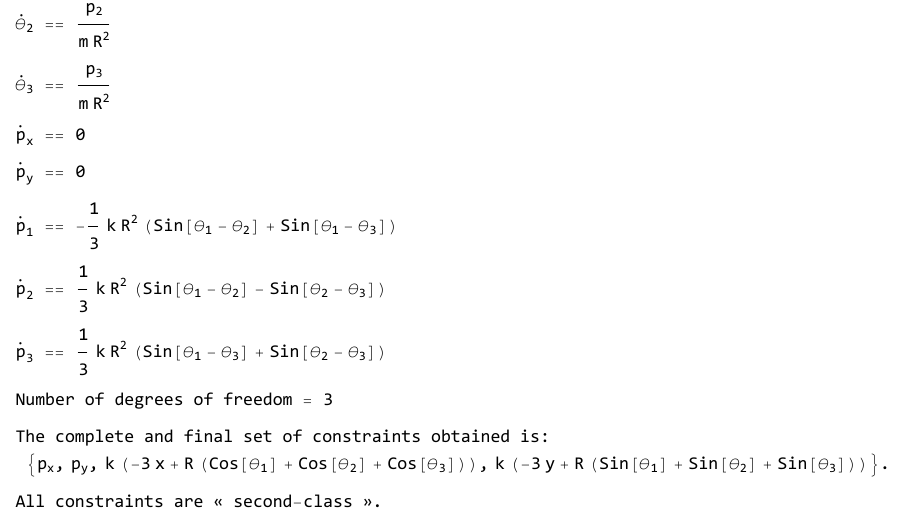}
    %\caption{Caption}
    %\label{fig:enter-label}
\end{figure}

%It is remarkable to describe briefly what \texttt{SystemInvoultionAnalysis} achieved. It obtained the secondary constraints and attached them new independent variables: $\varphi$ and $\rho$. Then after evolving the final set of constraints, and noticing that they are not in involution, calculated the equations of motion for every dynamic variable with the expected formula. All the results were shown in the subindex format as expected. It also exposed the number of degrees of freedom of the system and the classification of the constraints.

The execution of \texttt{SystemInvoultionAnalysis}  produces several key outcomes. It successfully identifies the secondary constraints and assigns them new independent variables: $\varphi$ and $\rho$. After evolving the final set of constraints and determining that they are not in involution, the function calculates the equations of motion for each dynamic variable using the expected formulas. All results are displayed in the subscript format as anticipated. Additionally, the function reveals the number of degrees of freedom in the system and classifies the constraints accordingly.

\section*{Acknowledgements}
The authors welcomes the support of the Universidad Juárez Autónoma de Tabasco for providing a suitable work environment while this research was carried out. J.M.C. also thanks the National Council of Human Studies, Science and Technology  (CONAHCYT) of México for their support through a grant for postdoctoral studies under Grant No. 3873825. We thank E. Chan-López for the discussions and insights regarding the work.

%\bibliography{sn-bibliography}

\end{document}